# Mathematical and numerical modelling of rapid transients at partially lifted sluice gates

Luca Cozzolino[1], Giada Varra[2], Luigi Cimorelli[3], Renata Della Morte[4]

**Abstract**

*The present paper deals with the modelling of rapid transients at partially lifted sluice gates from both a mathematical and numerical perspective in the context of the Shallow water Equations (SWE). First, an improved exact solution of the dam-break problem is presented, assuming (i) the dependence of the gate contraction coefficient on the upstream flow depth, and (ii) a physically congruent definition for the submerged flow equation. It is shown that a relevant solution always exists for any set of initial conditions, but there are also initial conditions for which the solution is multiple. In the last case, a novel disambiguation criterion based on the continuous dependence of the solution on the initial conditions is used to select the physically congruent one among the alternatives. Secondly, a one- (1-d) and a two-dimensional (2-d) form of a SWE Finite Volume numerical model - equipped with an approximate Riemann solver for the sluice gate treatment at cells interfaces – are presented. It is shown that the numerical implementation of classic steady state gate equations (classic equilibrium approach) leads to unsatisfactory numerical results in the case of fast transients, while a novel relaxed version of these equations (non-equilibrium approach) supplies very satisfactory results both in the 1-d and 2-d case. In particular, the 1-d numerical model is tested against (i) the proposed novel exact solutions and (ii) recent dam-break laboratory results. The 2-d model is verified by means of a test in a realistic detention basin for flood regulation, demonstrating that the novel findings can be promptly applied in real-world cases.*

Corresponding author: Giada Varra

E-mail address: giada.varra@uniparthenope.it

Address: Dipartimento di Ingegneria, Università degli Studi di Napoli Parthenope, Isola C4, 80143 Napoli (Italy)

## 1. Introduction

Sluice gates are commonly used as regulation structures in rivers and irrigation canals (Islam et al. 2008, van Thang et al. 2010), flow measurement devices in open channels (Silva and Rijo 2017, Kubrak et al. 2020), and control structures in flood detention basins (Morales-Hernández et al. 2013).

While sluice gates have been systematically studied considering steady flow conditions (Roth and Hager 1999, Lin et al. 2002, Defina and Susin 2003, Belaud et al. 2009, Habibzadeh et al. 2012, Bijankhan et al. 2012b, Castro-Orgaz et al. 2013), less attention has been paid to transient flows caused by rapid gate manoeuvres or interaction with waves propagating along the channel. There are, of course, some exceptions. De Marchi (1945) and Kubo and Shimura (1981) studied the negative wave generated upstream by the instantaneous partial uplift of a gate in a rectangular channel with water initially at rest, while Montuori and Greco (1973) studied the sudden manoeuvre (opening or closure) that caused the superposing of moving waves on preceding steady flow conditions. The exact solutions supplied by De Marchi (1945) and Kubo and Shumira (1981) were experimentally confirmed by Yamada (1992) and Reichstetter and Chanson (2013), while the exact solutions by Montuori and Greco (1973) were confirmed by their own laboratory experiments. Similar exact solutions are also present in classic textbooks such as those by Chow (1959) and Henderson (1966).

Among these studies, the work by Cozzolino et al. (2015) is particularly relevant here because a systematic analysis of the dam-break with partially lifted sluice gate in the context of the Shallow water Equations (SWE) was carried out by considering constant contraction coefficient with the adoption of the energy-momentum method by Henry (1950) for evaluating the discharge issuing under the gate in submerged flow conditions. Cozzolino et al. (2015) showed that there were initial conditions for which the dam-break problem exhibited multiple solutions and proposed a disambiguation criterion based on discharge maximization under the gate. In addition, they showed that there were initial conditions for which the dam-break problem exhibited no exact solution. The exact solutions by Cozzolino et al. (2015), which were subsequently verified for small gate opening by Monge-Gapper and Serrano-Pacheco (2021) using a smooth particle hydrodynamics model, are now a benchmark test for existing numerical models (Cui et al. 2019, Leakey et al. 2020, Delestre et al. 2023).

The lack of solution to the dam-break problem with partially lifted sluice gate for certain initial conditions is due to the choice of the gate equations made in Cozzolino et al. (2015). Despite a constant value of the gate contraction coefficient is commonly used in the technical literature (Lin et al. 2002, Jaafar and Merkley 2010, Wu and Rajaratnam 2015), theoretical studies (Cisotti 1908, Marchi 1953, Belaud et al. 2009), numerical computations (Montes 1997, Kim 2007, Lazzarin et al. 2023), and laboratory experiments (Rajaratnam and Subramanya 1967, Rajaratnam 1977, Defina and Susin 2003, Lazzarin et al. 2023), show that the contraction coefficient depends on the gate opening and the upstream flow depth. In addition, it is well known that the energy-momentum method by Henry (1950) is unable to calculate the discharge under the gate in the transitional region between free and submerged flow, causing the formation of a non-physical discontinuity in the gate discharge equation (Bijankhan et al. 2011, 2012a). This issue should be corrected by continuously connecting the free and submerged flow gate equations (Cunge et al. 1980).

The SWE model with sluice gate interior boundary conditions has been traditionally solved with the Method of Characteristics (Cunge et al. 1980, Islam et al. 2008), or locally coupling the Finite Volume method with the Method of Characteristics (Jaafar and Merkley 2010). The simultaneous solution of channel flow and gate equations with these approaches may lead to non-linear polynomial equations with order up to twelve (Ellis 1976), for which the existence of a solution is not granted. Recently, the weak coupling of sluice gate and channel flow equations through the fluxes that the structure exchanges with the channel flow has emerged as a viable alternative in Finite Volume schemes (Zhao et al. 1994). The computation of these fluxes has been often carried out by approximating the solution of a local sluice gate Riemann problem (Morales-Hernández et al. 2013, Lacasta et al. 2014, Cozzolino et al. 2015, Cui et al. 2019, Leakey et al. 2020). Nonetheless, this approach requires that the corresponding exact solutions are known in advance for benchmarking and constructing improved approximate Riemann solvers. Except for the numerical approach by

Cozzolino et al. (2015), current numerical methods do not recognize the existence of multiple Riemann solutions for certain initial conditions, and they lack a mechanism to cope with the solution multiplicity.

In the present paper, we construct novel SWE exact solutions of the dam-break at partially lifted sluice gates using variable contraction coefficient (Defina and Susin 2003) and the submerged flow gate equations by Bijankhan et al. (2012b). The novel solutions are improved with respect to those by Cozzolino et al. (2015) because the discharge gap in the transitional region between free and submerged flow is eliminated using viable experimental gate equations. We show that the dam-break solution always exists for any set of initial conditions, but there are certain initial conditions for which the solution is multiple. In this case, a criterion based on the continuous dependence of the solution on the initial conditions is used to pick up the relevant solution among the alternatives. In addition, we construct 1- and 2-d SWE Finite Volume models equipped with an approximate Riemann solver for the sluice gate treatment at cells interfaces. We show that the classic steady state gate equations lead to unsatisfactory results in the case of fast transients' numerical computation, and we propose a relaxed version of these equations, here called *non-equilibrium approach*, which coincides with the classic equations in the case of steady flow, and it is best suited for the construction of the approximate Riemann solver. The 1-d numerical model with the non-equilibrium approach for the gate treatment captures the novel exact dam-break solutions and the dam-break laboratory results by Lazzarin et al. (2023), while the 2-d numerical model is tested using a realistic 2-d detention basin for flood regulation.

The rest of the paper is organized as follows: in Section 2, the gate equations are presented; in Section 3, the exact solution of the dam-break problem with partially lifted sluice gate is constructed, and a novel disambiguation criterion is proposed; in Sections 4 and 5, 1-d and 2-d SWE models that incorporate the non-equilibrium numerical approach are described and tested; in Section 6, the novel disambiguation criterion is compared with the one by Lazzarin et al. (2023); finally, the paper is closed by a Conclusions section.

## 2. Sluice gate model

When fluid flows interact with a sluice gate, the corresponding regime is called orifice flow. In this case, two distinct flow conditions are possible, namely the free and the submerged flow (Henderson 1966). In free flow conditions (see Figure 1a), the supercritical jet issuing under the gate is open to the atmosphere. In contrast, in submerged flow conditions, the jet under the gate is overlaid by the downstream subcritical flow, which is characterised by intense turbulent motion (see Figure 1b). Finally, the regime where the flow free surface does not touch the gate lip and there is no interaction with the gate is referred to as non-orifice flow (Figure 1c).

In the present Section, the gate equations are presented, and a bifurcation phenomenon is introduced.

### 2.1. Sluice gate equations

In steady free flow conditions, the unit-width discharge $q_F$ under the gate depends on the gate opening $a$ and the upstream depth $h_u$, which is measured at a distance from the gate sufficient to re-establish gradually varied flow (Figure 1a). The cross-section where the depth of the supercritical jet issuing under the gate is minimum and the flow is gradually varied is called *vena contracta*.

At the generic cross-section, the energetic content of the flow is measured by the total head $H = h + q^2/(2gh^2)$, where $h$ is the flow depth and $q$ is the unit-width discharge. If the energy loss through the gate is neglected and steady state conditions are assumed, the invariance of total head and unit-width discharge between the upstream cross-section and the *vena contracta* implies

(1) $\quad h_u + \dfrac{q_F^2}{2gh_u^2} = h_c + \dfrac{q_F^2}{2gh_c^2},$

where $h_c$ is the *vena contracta* flow depth. After solving for $q_F$ and rearranging, the free flow gate equation can be written as (Rouse 1946, Henderson 1966, Defina and Susin 2003)

$$(2) \quad q_F = \frac{C_c a}{\sqrt{1+\dfrac{C_c a}{h_u}}} \sqrt{2 g h_u},$$

where $C_c = h_c/a$ is the contraction coefficient, i.e., the ratio between the *vena contracta* flow depth and the gate opening. From the preceding, we observe that the use of Eq. (2) implies that the value $H_{u,req} = h_u \left[1 + (C_c a/h_u)^2 / (1 + C_c a/h_u)\right]$ of the upstream head $H_u$ is required to make the discharge $q_F$ pass under the gate in free flow conditions when the upstream flow depth is $h_u$. The head $H_{u,req}$ is obviously greater than $h_u$.

In the present paper, the parametric formulation by Defina and Susin (2003)

$$(3) \quad \begin{aligned} r(\theta) &= 0.153\theta^2 - 0.451\theta + 0.727 \\ C_c &= 1 - r(\theta)\sin\theta \\ a/h_u &= 1 - r(\theta)[1-\cos\theta] \end{aligned}$$

is used to establish a relationship between $C_c$ and the relative opening $a/h_u$. In Eq. (3), which is based on numerous experimental data, the parameter $\theta$ falls in the range $[0, 2.499[$.

The depth $h_c^\#$ of the subcritical flow conjugated by a hydraulic jump to the supercritical flow at the *vena contracta* (Figure 1a) is given by

$$(4) \quad h_c^\# = \frac{C_c a}{2}\left(-1 + \sqrt{1 + F_c^2}\right),$$

where $F_c^2 = q_F^2 / \left[ g(C_c a)^3 \right]$ is the squared Froude number associated to the *vena contracta* supercritical flow.

In steady submerged flow conditions (Figure 1b), the unit-width discharge $q_S$ under the gate also depends on the tailwater depth $h_t$, i.e., the depth of the downstream subcritical flow measured at a distance sufficient to re-establish the gradually varied flow (Henry 1950, Rajaratnam and Subramanya 1967, Lozano et al. 2009). In the present paper, $q_S$ is calculated with the experimental expression by Bijankhan et al. (2012b)

$$(5) \quad q_S = q_F \left( \frac{\dfrac{h_u - h_t}{a}}{\alpha \left[ \dfrac{h_u - h_c^\#}{a} \right]^\beta + \dfrac{h_u - h_t}{a}} \right)^{\frac{3}{2}\eta},$$

where $\alpha = 2.01$, $\beta = 0.921$, and $\eta = 0.2848$, are interpolation parameters, while $q_F$, $C_c$, and $h_c^\#$, are computed using Eqs. (2)-(4), respectively. Free flow conditions are possible only when $h_t < h_c^\#$, while submerged flow conditions are established for $h_c^\# < h_t \leq h_u$. The value $h_t = h_c^\#$ of the tailwater depth represents the limit condition between free and submerged flow (Rajaratnam and Subramanya 1967, Lin et al. 2002, Habibzadeh et al. 2011).

Since the term inside the parentheses to the right-hand side of Eq. (5) is minor than one, it follows that $q_S \leq q_F$ for given $a$ and $h_u$. In other words, a tailwater depth $h_t > h_c^\#$ reduces the discharge issuing under the gate with respect to the free flow case. The formulation of Eq. (5) is chosen here because it is based on numerous experimental data and satisfies the congruency conditions $q_S = q_F$ for $h_t = h_c^\#$ and $q_S = 0$ for $h_t = h_u$, and it continuously connects the submerged and the free flow gate equations.

[Insert Figure 1 about here]

## 2.2. Free flow multiple solutions with variable contraction coefficient

Before considering the dam-break problem solution with partially lifted sluice gate, it is instructive to contrast the effect of variable and constant $C_c$ on the discharge calculated with Eq. (2). In addition to the experimental $C_c$ of Eq. (3) by Defina and Susin (2003), the theoretical formulation by Marchi (1953), which is based on the assumption of irrotational flow with gravity effects included, will be considered in the present Section.

In Figure 2a, the experimental contraction coefficient $C_c$ by Defina and Susin (2003) is plotted as a function of the relative opening $a/h_u$ (thick black line), together with the theoretical $C_c$ by Marchi (1953) (dashed black line) and the constant value $C_c = 0.611$ (thin black line) usually applied in the literature (Lin et al. 2002, Jaafar and Merkley 2010, Cozzolino et al. 2015). The inspection of the figure shows that the theoretical expression of $C_c$ by Marchi (1953) is a convex function of $a/h_u$ that exhibits a minimum in $a/h_u = 0.29$ and satisfies the condition $C_c = 1.0$ for $a/h_u = 1.0$. The last condition expresses the fact that no flow contraction is expected when the gate lip barely trims the flow free surface. Similarly, the experimental $C_c$ by Defina and Susin (2003) is convex with a minimum in $a/h_u = 0.48$ and satisfies the no flow-contraction condition for $a/h_u = 1.0$. Trivially, this requirement cannot be met by a constant value of $C_c$.

The shape exhibited by different $C_c$ models has an influence on the discharge issued under the gate in free flow conditions, at least for high values of the relative opening $a/h_u$. Let $F_F^2 = q_F^2/(ga^3)$ be a squared gate Froude number where $q_F$ is computed by means of Eq. (2). By definition, $F_F^2$ is representative of the unit-width discharge issuing under the sluice gate for given opening $a$ in free flow conditions. In Figure 2b, $F_F^2$ is plotted as a function of $a/h_u$ using the $C_c$ expression by Defina and Susin (2003) (thick black line), together with the theoretical $C_c$ by Marchi (1953) (dashed black line) and the constant value $C_c = 0.611$ (thin black line).

The inspection of Figure 2b shows that the different definitions of $C_c$ lead to similar values of $F_F^2$ for $a/h_u < 0.7$, and this explains why the use of the constant $C_c = 0.611$ is widespread in practical applications and literature. Nonetheless, significant differences are evident for higher values of the relative opening $a/h_u$. The function $F_F^2$ with constant $C_c$ is strictly decreasing in the entire interval $a/h_u \in\, ]0, 1]$, implying that a single value of $a/h_u$ is associated to each value of $F_F^2$. On the contrary, the functions $F_F^2$ computed with the theoretical formula by Marchi (1953) and the experimental expression of Eq. (3) by Defina and Susin (2003) exhibit a minimum in $(a/h_u)_{lim} = 0.83$ and $(a/h_u)_{lim} = 0.86$, respectively. This implies that there are values of $F_F^2$ that can be associated to two distinct values of $a/h_u$ when an expression that satisfies the no flow-contraction condition is used for $C_c$. For the same curves we observe from Figure 2b that, congruently with the physical intuition, $q_F$ is an increasing function of $h_u$ for $a/h_u < (a/h_u)_{lim}$. Vice versa, $q_F$ is a decreasing function of $h_u$ for $a/h_u > (a/h_u)_{lim}$. Lazzarin et al. (2023) have associated this behaviour with flow instability phenomena at gates with high relative opening.

In the following, it will be shown that the shape of the $C_c$ curve has a dramatic influence on the existence and uniqueness of the dam-break problem solution, even in the simplest case of dam-break on dry bed.

[Insert Figure 2 about here]

## 3. Exact solution to the dam-break problem at partially lifted gates

Under the assumptions of a constant-width rectangular channel with a horizontal frictionless bed, the 1-d Shallow water Equations can be written as (Toro 2001, LeVeque 2002)

(6) $\dfrac{\partial \mathbf{u}}{\partial t} + \dfrac{\partial \mathbf{f}(\mathbf{u})}{\partial x} = 0$.

In Eq. (6), the meaning of the symbols is as follows: $x$ is the longitudinal coordinate; $t$ is the time variable; $\mathbf{u}(x,t) = (h \quad hu)^T$ is the vector of the conserved variables, where $h(x,t)$ is the flow depth, $u(x,t)$ is the velocity, and $T$ is the matrix transpose symbol; finally, $\mathbf{f}(\mathbf{u}) = (hu \quad 0.5gh^2 + hu^2)^T$ is the flux vector, where $g = 9.81$ m/s² is the gravity acceleration.

The Riemann problem is the initial value problem where Eq. (6) is solved with the discontinuous initial conditions

$$(7) \quad \mathbf{u}(x,0) = \begin{cases} \mathbf{u}_L, & x < 0 \\ \mathbf{u}_R, & x > 0 \end{cases},$$

where $\mathbf{u}_L = (h_L \quad h_L u_L)^T$ and $\mathbf{u}_R = (h_R \quad h_R u_R)^T$ are the left and right initial states. The solution to the Riemann problem is self-similar in the plane $(x, t)$, i.e., it exists a vector $\mathbf{w}(x/t)$ such that $\mathbf{u}(x,t) = \mathbf{w}(x/t)$ for $t > 0$, and it consists of a sequence of constant states connected by moving waves (shocks or rarefactions) and possibly by a standing wave in $x = 0$ that models the action exerted by the gate on the flow.

In the $(h,u)$ plane, the curve of the states $\mathbf{u}$ connected to the reference state $\mathbf{u}_0 = (h_0 \quad h_0 u_0)^T$ by a rarefaction contained into the $i$-th characteristic field has equation (Toro 2001, LeVeque 2002, Han and Warnecke 2014)

$$(8) \quad u = f_{R,i}(h) = u_0 \mp 2\left(\sqrt{gh} - \sqrt{gh_0}\right).$$

From Eq. (8), it follows that the unit-width discharge corresponding to a state $\mathbf{u}$ along the rarefaction contained into the $i$-th characteristic field has equation (LeVeque 2002)

(9) $hu = q_{R,i} = hu_0 \mp 2h\left(\sqrt{gh} - \sqrt{gh_0}\right)$.

In Eqs. (8)-(9), the minus and the plus signs are related to the first and second characteristic field, respectively. The rarefaction is called direct and it is denoted with the symbol $R_i(\mathbf{u}_0)$ if $\mathbf{u}$ follows $\mathbf{u}_0$ along the $x$ axis in the Riemann self-similar solution, otherwise it is called backward and it is denoted with the symbol $R_i^B(\mathbf{u}_0)$. It is immediate to see that $u = f_{R,i}(h)$ is a strictly decreasing [increasing] function of $h$ for $i = 1$ [$i = 2$] (Han and Warnecke 2014). It is immediate to observe that, in the first characteristic field, the discharge $q_{R,i}$ is a decreasing function of $h$ for $h > \left(u_0 + 2\sqrt{gh_0}\right)^2 / (9g)$. This fact will be useful when the solution of the dam-break problem will be considered in Section 3.2.

Similarly, the curve of the ($h$, $u$) plane consisting of the states $\mathbf{u}$ connected to the reference state $\mathbf{u}_0$ by a shock contained into the $i$-th characteristic field has equation (Toro 2001, LeVeque 2002, Han and Warnecke 2014)

(10) $u = f_{S,i}(h) = u_0 \mp (h - h_0)\sqrt{\dfrac{g}{2}\left(\dfrac{1}{h} + \dfrac{1}{h_0}\right)}$,

and the celerity of the discontinuity separating $\mathbf{u}$ and $\mathbf{u}_0$ is

(11) $\sigma = u_0 \mp \sqrt{gh_0\left(\dfrac{h}{h_0} + 1\right)}$.

In Eqs. (10)-(11), the minus and the plus signs are related to the first and second characteristic field, respectively. The shock is called direct and it is denoted with the symbol $S_i(\mathbf{u}_0)$ if $\mathbf{u}$ follows $\mathbf{u}_0$ along the $x$ axis in the Riemann self-similar solution, otherwise it is called backward and it is denoted with the symbol $S_i^B(\mathbf{u}_0)$. Similar to what happens for rarefactions, $u = f_{S,i}(h)$ is a strictly decreasing [increasing] function of $h$ for $i = 1$ [$i = 2$] (Han and Warnecke 2014). Generalizing the use made in Eq. (4), the hash (#) superscript is used here to denote the subcritical state $\mathbf{u}_0^\#$ connected to the supercritical state $\mathbf{u}_0$ by means of a hydraulic jump, i.e., by a shock with null celerity. By definition, the unit-width discharges corresponding to $\mathbf{u}_0^\#$ and $\mathbf{u}_0$ coincide.

Let $\mathbf{u}_1 = (h_1 \quad h_1 u_1)^T = \mathbf{u}(0^-, t)$ and $\mathbf{u}_2 = (h_2 \quad h_2 u_2)^T = \mathbf{u}(0^+, t)$ be the states of the Riemann solution immediately to the left and to the right of the gate location $x = 0$, respectively. The self-similarity of the solution requires that $\mathbf{u}(0^-, t) = \mathbf{w}(0^-)$ and $\mathbf{u}(0^+, t) = \mathbf{w}(0^+)$ for $t > 0$, and this implies that $\mathbf{u}_1$ and $\mathbf{u}_2$ are constant in time. When non-orifice flow conditions are established, $\mathbf{u}_1$ and $\mathbf{u}_2$ either coincide because the Riemann solution is continuous through $x = 0$ or they differ because they are connected by a shock with null celerity. When free flow conditions are established, the states $\mathbf{u}_1$ and $\mathbf{u}_2$ are connected by the gate relation of Eq. (2) and the standing wave in $x = 0$ is denoted with the symbol $SW_f$. In case of submerged flow, the states $\mathbf{u}_1$ and $\mathbf{u}_2$ are connected by the gate relations of Eq. (5) and the standing wave is denoted with the symbol $SW_s$. Independent on the flow regime established (orifice or non-orifice), the mass conservation principle requires that the unit-width discharges $h_1 u_1$ and $h_2 u_2$ coincide. In turn, these discharges coincide with the discharge issuing under the gate in the case of orifice flow regime.

In the dam-break problem, the flow velocity is initially null, and the initial states reduce to $\mathbf{u}_L = (h_L \quad 0)^T$ and $\mathbf{u}_R = (h_R \quad 0)^T$. This corresponds to the situation where two reservoirs with water initially at rest are separated by a sluice gate in $x = 0$ that is suddenly lifted leaving an opening of height $a$. The dam-break solution is trivial when the gate is lifted enough to avoid subsequent interaction between the flow free-surface and the gate lip because the corresponding non-orifice regime is equivalent to the case of gate complete removal already discussed in the literature (Stoker 1957, Toro 2001, LeVeque 2002). In the present section, we generalize the dam break to consider the case where the flow interacts with the gate because the device is only partially lifted.

Without loss of generality, we will assume in the following that $h_L \geq h_R$, implying that the flow moves from the left to the right under the gate during the transient caused by the gate lifting. It follows that, in free flow conditions, the state $\mathbf{u}_2$ coincides with the *vena contracta* state $\mathbf{u}_c = (h_c \quad q_F)^T$, where $q_F$ is calculated using Eq. (2) with $h_u = h_1$. In submerged flow conditions, the state $\mathbf{u}_2$ is such that $\mathbf{u}_2 = (h_2 \quad q_S)^T$, where $q_S$ is calculated using Eq. (5) with $h_t = h_2$ and $h_u = h_1$. From the discussion related to Eq. (5) (Section 2.1), it follows that the limit tailwater state $\mathbf{u}_2$ in submerged flow conditions is $\mathbf{u}_c^{\#} = (h_c^{\#} \quad q_F)^T$, where $h_c^{\#}$ is computed using Eq. (4).

### 3.1 Preliminaries: dam-break on a dry bed and multiple solutions

To introduce the issue of multiple solutions to the dam-break problem at partially lifted gates, first we explore the solution of the dam-break on dry bed ($h_R = 0$ m), considering the parameters $a = 0.47$ m and $h_L = 1$ m (Test E1 of Table 1), with the $C_c$ definition of Eq. (3). In the following, we show that three distinct solutions are possible, as depicted in Figure 3. In all the solutions, the gate lifting causes the formation of a rarefaction wave $R_1(\mathbf{u}_L)$ that empties the left reach of the channel making the flow move from left to right.

In the first dam-break solution (Figure 3a), which coincides with the classic solution by Ritter (1892), a non-orifice condition is established because the flow free-surface corresponding to the state $\mathbf{u}_1$ does not touch the gate lip. In this case, the rarefaction $R_1(\mathbf{u}_L)$ directly connects the state $\mathbf{u}_L$ to the dry bed. Since the theory by Ritter (1892) supplies $h_1 = h_2 = 4h_L/9$ in $x = 0$, this type of solution is feasible when the initial relative opening $a/h_L$ satisfies the necessary condition $a/h_L > 4/9$. In the example considered, $h_1 = h_2 = 0.444$ m is obtained.

The second and the third solution (Figures 3b,c) are characterized by orifice free flow conditions, which are fully determined if the corresponding state $\mathbf{u}_1$ is known. To find $\mathbf{u}_1$, we introduce the state $\mathbf{u}_F = \begin{pmatrix} h_F & q_F \end{pmatrix}^T$ connected to $\mathbf{u}_L$ by the $R_1(\mathbf{u}_L)$ rarefaction and such that the corresponding unit-width discharge $q_{R,1}$ of Eq. (9) coincides with the free flow discharge $q_F$ of Eq. (2) where $h_u = h_F$. It follows that the flow depth $h_F$ corresponding to $\mathbf{u}_F$ satisfies the equation

$$(12) \quad 2\left(\sqrt{gh_L} - \sqrt{gh_F}\right)h_F = \frac{C_c a}{\sqrt{1 + \frac{C_c a}{h_F}}}\sqrt{2gh_F} \;.$$

Depending on the initial relative opening $a/h_L$, Eq. (12) admits one, two, or no solutions. Correspondingly, one, two, or no states $\mathbf{u}_1 = \mathbf{u}_F$ can be associated to orifice free flow conditions. When Eq. (12) exhibits two solutions, a subscript $l$ [$h$] characterizes the lower [higher] value $h_{F,l}$ [$h_{F,h}$] of the flow depth solution and the corresponding state $\mathbf{u}_{F,l}$ [$\mathbf{u}_{F,h}$], determining a *lower* [*higher*] *solution*. In the example with $a = 0.47$ m and $h_L = 1$ m (Test E1 of Table 1), $h_{F,l} = 0.475$ m (Figure 3b) and $h_{F,h} = 0.609$ m (Figure 3c) are found.

From the preceding discussion, it follows that three distinct dam-break solutions on dry bed are possible for the example considered, namely one non-orifice and two orifice flow solutions (Figure 3). If the same procedure is repeated for different values of $a$ and $h_L$, the diagram of the relative depth $h_1/h_L$ as a function of the initial relative opening $a/h_L$ is obtained (see Figure 4). The inspection of the figure shows that the dam-break problem on dry bed admits a single orifice free flow solution for low initial relative openings ($a/h_L < 4/9$), while a single non-orifice flow solution is obtained at high values of the initial relative opening ($a/h_L > 0.495$). Finally, three distinct solutions to the dam-break problem on dry bed are possible for intermediate values of the initial relative opening ($a/h_L \in [4/9, 0.495]$).

In the next Sections, we will construct the general solution to the dam-break problem at partially lifted gates for the case with $h_R > 0$ and we will provide a criterion for the disambiguation of multiple solutions.

[Insert Figure 3 about here]

[Insert Figure 4 about here]

[Insert Table 1 about here]

**3.2 General solution to the dam-break problem**

The graphic solution to the classic dam-break problem with completely open gate and $h_L \geq h_R$ is trivially found in the plane ($h,u$) by determining the intersection $\mathbf{u}_M$ between the direct rarefaction curve $R_1(\mathbf{u}_L)$ and the backward shock curve $S_2^B(\mathbf{u}_R)$ (Stoker 1957, LeVeque 2002). The solution of this dam-break always exists for $h_L \geq h_R$ and it is also unique because the curves of the ($h, u$) plane corresponding to $R_1(\mathbf{u}_L)$ and $S_2^B(\mathbf{u}_R)$ are strictly decreasing and strictly increasing, respectively (see Eqs. [8] and [10]).

The direct application of this procedure is not meaningful in the case of partially lifted gate because the waves $R_1(\mathbf{u}_L)$ and $S_2^B(\mathbf{u}_R)$ do not take into account the presence of the device. Following Marchesin and Paes-Leme (1986) and Han and Warnecke (2014), it is possible to construct a new curve of the ($h, u$) plane, called L-M, which generalizes $R_1(\mathbf{u}_L)$ and incorporates the effects introduced by the gate. The intersection $\mathbf{u}_M$ of $S_2^B(\mathbf{u}_R)$ with a branch of the curve L-M individuates the solution wave configuration.

In the following, we will separately consider the cases of low ($a/h_L < 4/9$), high ($a/h_L > 0.495$), and intermediate initial relative opening ($a/h_L \in [4/9, 0.495]$).

### 3.2.1 Low initial relative opening

For low initial relative openings ($a/h_L < 4/9$), the state $\mathbf{u}_L$ cannot be connected by the rarefaction $R_1(\mathbf{u}_L)$ to a non-orifice state u1 characterized by $h_1 < a$ (see Section 3.1), there is only one solution to Eq. (12) (see Section 3.1). For this reason, it is possible to enumerate the three distinct wave configurations represented in Figure 5, corresponding to the exact solutions of the dam-break tests from E2 to E4 contained in Table 1, which are characterized by $a/h_L = 0.2$ and increasing values of $h_R$.

In the wave configurations of Figures 5a and 5b (Tests E2 and E3 of Table 1, respectively), $R_1(\mathbf{u}_L)$ connects the state $\mathbf{u}_L$ to the state $\mathbf{u}_1 = \mathbf{u}_F$ while the $SW_f$ wave connects $\mathbf{u}_1$ to $\mathbf{u}_2$. Being $a/h_L < 4/9$, the orifice free flow conditions determine a unique state $\mathbf{u}_1 = \mathbf{u}_F$ for given $\mathbf{u}_L$ and $a$ (see Section 3.1), and this in turns determines a unique state $\mathbf{u}_2 = \mathbf{u}_c$. Finally, a $R_1(\mathbf{u}_2)$ wave (Figure 5a) or a $S_1(\mathbf{u}_2)$ wave (Figure 5b) connects the state $\mathbf{u}_{mid} = (h_{mid} \quad h_{mid} u_{mid})^T$ to $\mathbf{u}_2$, while $S_2^B(\mathbf{u}_R)$ connects $\mathbf{u}_{mid}$ to $\mathbf{u}_R$.

In the wave configuration of Figure 5c (Test E4 of Table 1), the $R_1(\mathbf{u}_L)$ rarefaction connects the states $\mathbf{u}_L$ and $\mathbf{u}_1$ while the $SW_s$ wave connects $\mathbf{u}_1$ to the tailwater state $\mathbf{u}_2$. Finally, the $S_2^B(\mathbf{u}_R)$ shock connects the state $\mathbf{u}_2$ and the state $\mathbf{u}_R$. Recalling that $q_S \leq q_F$, it follows that $h_1 \geq h_F$ (see comments to Eq. [9]), which in turn implies that the state $\mathbf{u}_1$ lies on $R_1(\mathbf{u}_L)$ between $\mathbf{u}_L$ and $\mathbf{u}_F$. To enable the solution of the dam-break problem, we introduce an additional curve of the plane $(h, u)$, called Tailwater curve (TC), which is defined as the locus of the tailwater states $\mathbf{u}_2$ associated through the $SW_s$ wave to each state $\mathbf{u}$ along the $R_1(\mathbf{u}_L)$ curve between $\mathbf{u}_L$ and $\mathbf{u}_F$.

The preceding observations suggest the construction of the L-M curve as the union of the following three branches (Figure 6a):

- the locus of the states $\mathbf{u}_{mid}$ connected to $\mathbf{u}_2 = \mathbf{u}_c$ by a $R_1(\mathbf{u}_2)$ rarefaction entirely developing to the right of the gate;
- the locus of the states $\mathbf{u}_{mid}$ connected to $\mathbf{u}_2 = \mathbf{u}_c$ by a $S_1(\mathbf{u}_2)$ shock entirely developing to the right of the gate;
- the locus of the states $\mathbf{u}_2$ along TC.

In Figure 6a, the L-M curve is plotted in the $(h, u)$ plane for the initial relative opening $a/h_L = 0.2$. Trivially, the state $\mathbf{u}_c$ represents the transition between the first and the second branch of L-M, while $\mathbf{u}_c^{\#}$ represents the transition between the second and the third branch.

In Figures 6b,c,d, we consider the graphical solution of the dam-break tests from E2 to E4 of Table 1. For Test E2 (Figure 6b), the intersection $\mathbf{u}_M$ between the L-M and $S_2^B(\mathbf{u}_R)$ curves lies on $R_1(\mathbf{u}_2)$, implying that an orifice free flow is established and that the state $\mathbf{u}_{mid} = \mathbf{u}_M$ is separated from the state $\mathbf{u}_2$ by a rarefaction. As already mentioned, the corresponding flow depth exact solution at time $t = 5$ s is represented in Figure 5a. For Test E3 (Figures 5b, 6c), the state $\mathbf{u}_{mid} = \mathbf{u}_M$ is separated from the state $\mathbf{u}_2$ by a shock because $\mathbf{u}_M$ lies on $S_1(\mathbf{u}_2)$. Finally, Test E4 (Figures 5c, 6d) differs from the preceding examples because the intersection $\mathbf{u}_M$ lies on the TC curve. This implies that the flow is submerged and $\mathbf{u}_2 = \mathbf{u}_M$ is tailwater state.

As a final consideration, we observe from Figure 6a that the L-M curve is continuous and strictly decreasing, while $S_2^B(\mathbf{u}_R)$ is continuous and strictly increasing. This proves that the solution of the dam-break problem always exists and it is unique in case of low initial relative opening.

[Insert Figure 5 about here]
[Insert Figure 6 about here]

*3.2.2 High initial relative opening*

For high initial relative openings ($a/h_L > 0.495$), the orifice free flow is forbidden because Eq. (12) has no solution (see Section 3.1). In this case, it is possible to enumerate the two wave configurations represented in Figure 7, which correspond to dam-break examples with $a/h_L = 0.6$ and increasing values of $h_R$.

In the wave configuration of Figure 7a (Test E5 of Table 1), a non-orifice flow regime is established where the $R_1(\mathbf{u}_L)$ and $S_2^B(\mathbf{u}_R)$ waves are separated by the intermediate state $\mathbf{u}_{mid}$, with $h_{mid} \leq a$. Having defined $\mathbf{u}_a = (h_a \quad h_a u_a)^T$ as the state along $R_1(\mathbf{u}_L)$ which satisfies the condition $h_a = a$, this implies that $\mathbf{u}_{mid}$ must lie along $R_1(\mathbf{u}_L)$ between the states $\mathbf{u}_a$ and the dry bed state.

The wave configuration of Figure 7b (Test E6 of Table 1) coincides with that of Figure 5c, and it is characterized by orifice submerged flow conditions. The construction of the TC curve is accomplished by using the $SW_s$ wave to associate a state $\mathbf{u}_2$ to each state $\mathbf{u}$ along the $R_1(\mathbf{u}_L)$ curve between $\mathbf{u}_L$ and $\mathbf{u}_a$. The state $\mathbf{u}_a$ itself lies on the TC curve because the case $\mathbf{u}_1 = \mathbf{u}_a$ coincides with the condition of gate lip barely trimming the free surface, which implies no-contraction ($h_2 = h_1$) and then $\mathbf{u}_2 = \mathbf{u}_a$.

From the preceding, it follows that the L-M curve consists of the following two branches (Figure 8a):

- the locus of the states $\mathbf{u}_{mid}$ along $R_1(\mathbf{u}_L)$ between the state $\mathbf{u}_a$ and the dry bed state;
- the locus of the states $\mathbf{u}_2$ along TC.

From Figure 8a, we observe that the L-M curve corresponding to the dam-break with high relative opening is continuous and strictly decreasing, proving that the solution of the dam-break problem always exists and it is unique in this case. In Figures 8b and 8c, we consider the graphical solution of the dam-break tests E5 and E6 (Table 1), respectively, characterized by high initial relative opening $a/h_L = 0.6$. In the case of Test E5 (Figure 8b), the intersection $\mathbf{u}_M$ between the L-M and $S_2^B(\mathbf{u}_R)$ curves lies on $R_1(\mathbf{u}_L)$, implying that a non-orifice regime is established. The corresponding exact solution (flow depth) at time $t = 5$ s is represented in Figure 7a. In the case of Test E6 (Figures 7b, 8c), the intersection $\mathbf{u}_M$ lies on the TC curve, implying that the flow is submerged with state $\mathbf{u}_2 = \mathbf{u}_M$ coinciding with the tailwater state.

[Insert Figure 7 about here]

[Insert Figure 8 about here]

*3.2.3 Intermediate initial relative opening*

For intermediate values of the initial relative opening ($a/h_L \in [4/9, 0.495]$), it is possible to consider three different L-M curves, each connected to a corresponding solution of the dam-break on dry bed (Section 3.1), namely one non-orifice flow solution and two distinct orifice-flow solutions. In the following, we will find that only one of these L-M curves is continuous, i..e., only one of the L-M curves ensures the existence of a dam-break solution for every possible right flow depth $h_R$.

The first L-M curve considered is the one where the state $\mathbf{u}_1$ in the orifice free-flow solutions coincides with $\mathbf{u}_{F,h}$. In Figure 9, the corresponding flow depth exact solution at $t = 5$ s for the dam-break tests from E7 to E9 (Table 1) with $a/h_L = 0.47$ is represented. In Figures 9a and 9b, $R_1(\mathbf{u}_L)$ connects the state $\mathbf{u}_L$ to the state $\mathbf{u}_1 = \mathbf{u}_{F,h}$ while the wave $SW_f$ connects $\mathbf{u}_1$ to the corresponding $\mathbf{u}_2 = \mathbf{u}_c$. Finally, a $R_1(\mathbf{u}_2)$ wave (Figure 9a) or a $S_1(\mathbf{u}_2)$ wave (Figure 9b) connects the state $\mathbf{u}_{mid}$ to $\mathbf{u}_2$, while $S_2^B(\mathbf{u}_R)$ connects $\mathbf{u}_{mid}$ to $\mathbf{u}_R$. In the wave configuration of Figure 9c, the $R_1(\mathbf{u}_L)$ rarefaction connects $\mathbf{u}_L$ and $\mathbf{u}_1$, while the submerged flow $SW_s$ connects $\mathbf{u}_1$ to the tailwater state $\mathbf{u}_2$, and the $S_2^B(\mathbf{u}_R)$ shock connects $\mathbf{u}_2$ and $\mathbf{u}_R$.

The inspection of Figure 9 shows that the dam-break wave configurations connected to the higher solution $\mathbf{u}_{F,h}$ coincide with the wave configurations obtained for low values of $a/h_L$ (Figure 5).

Following the methods of Section 3.2.1, it is possible to construct the corresponding L-M curve (Figure 10a). This curve is continuous and strictly decreasing, which proves that the corresponding dam-break solution always exists and it is unique. In Figures 10b,c,d, the graphical solution of the dam-break tests from E7 to E9 (Table 1) is represented.

[Insert Figure 9 about here]

[Insert Figure 10 about here]

The wave configurations for the case $\mathbf{u}_1 = \mathbf{u}_{F,l}$ (lower solution of Eq. [12]) coincide with those of Figures 5 and 9, and they are not reported here for the sake of brevity. For this reason, the corresponding L-M curve (Figure 11a) can be constructed following the methods contained in Section 3.2.1. Similarly, the wave configurations related to the non-orifice solution of the dam-break on dry bed coincide with those of Figure 7, and the corresponding L-M curve can be constructed following the methods from Section 3.2.2 (Figure 11b). The inspection of Figure 11 shows that these two additional L-M curves valid for $a/h_L = 0.47$ are discontinuous because they exhibit a solution gap along the TC curve, implying that there are states $\mathbf{u}_R$ for which the $S_2^B(\mathbf{u}_R)$ curve does not intersect L-M. In other words, there are initial conditions $\mathbf{u}_R$ for which there is no mathematical solution to the dam-break if one the L-M curves of Figure 11 is used.

To shed light on the origin of the gap on these L-M curves, we observe that the states $\mathbf{u}_2$ along the TC curve satisfy the equalities $h_2 u_2 = q_{R,1}$ and $h_2 u_2 = q_S$, which is possible only when $q_{R,1} \leq q_F$ (see comments to Eq. [5]). In Figure 12, the unit-width discharges $q_{R,1}$ and $q_F$ are plotted as a function of $h_1$ for the case of $h_L = 1$ m and $a = 0.47$ m. The inspection of the figure shows that the condition $q_{R,1} \leq q_F$ is satisfied in the entire interval of $h_1$ values used for the construction of the TC curve related to $\mathbf{u}_{F,h}$ (Figure 12a), while this is not true in the case of the L-M curves related to $\mathbf{u}_{F,l}$ (Figure 12b) and to the non-orifice solution of the dam-break on-dry bed (Figure 12c).

[Insert Figure 11 about here]

[Insert Figure 12 about here]

### 3.3 A disambiguation criterion for multiple solutions

The discussion of Section 3.2 shows that the exact solution to the dam-break at partially lifted gate can be obtained by intersecting the L-M curve, which depends on the state $\mathbf{u}_L$ and the opening $a$, with the $S_2^B(\mathbf{u}_R)$ curve.

When the initial relative opening is such that $a/h_L < 4/9$ or $a/h_L > 0.495$, the L-M curve is strictly decreasing and the intersection with the $S_2^B(\mathbf{u}_R)$ curve exists and it is unique for any value of the initial downstream depth $h_R$. In other words, the solution to the dam-break problem always exists and it is unique for $a/h_L < 4/9$ and $a/h_L > 0.495$.

When $a/h_L \in [4/9, 0.495]$, it is possible to plot three distinct L-M curves, implying that there are up to three potential solutions to the dam-break problem. While the L-M curve connected to the higher solution $\mathbf{u}_{F,h}$ of Eq. (12) is continuous and strictly decreasing, implying that the dam-break solution exists and it is unique for any value of $h_R$ if this L-M curve is used, the two additional L-M curves corresponding to $a/h_L \in [4/9, 0.495]$ do not even ensure the existence of a solution for certain values of $h_R$ because they are characterized by a solution gap.

The preceding observations lead to the following

**Proposition 1**. The exact solution to the dam-break problem with partially lifted gate always exists and it is unique. In the interval $a/h_L \in [4/9, 0.495]$, the relevant L-M curve is the one connected to the higher solution $\mathbf{u}_{F,h}$ of Eq. (12), while the two remaining L-M curves must be discarded.

The choice made for disambiguating multiple solutions of the dam-break problem with partially lifted gate ensures the internal congruency of the mathematical model in every circumstance.

Following Proposition 1, the solution to the dam-break problem on dry bed corresponds to a orifice flow regime when $a/h_L \in \,]0, 0.495]$, while non-orifice flow regime is obtained for $a/h_L > 0.495$.

## 4. One-dimensional numerical modelling

If the friction is added to Eq. (6), the 1-d SWE model in a horizontal rectangular channel with uniform width $B$ can be rewritten as

$$(13) \quad \frac{\partial \mathbf{u}}{\partial t} + \frac{\partial \mathbf{f}(\mathbf{u})}{\partial x} = -\mathbf{S}_f(\mathbf{u}).$$

where

$$(14) \quad \mathbf{S}_f(\mathbf{u}) = \begin{pmatrix} 0 & ghS_{f,x} \end{pmatrix}^T$$

is the friction vector. In the present work, the friction slope $S_{fx}$ of Eq. (14) is computed by means of the Manning's formula

$$(15) \quad S_{f,x} = \frac{n_M^2 u |u|}{R^{4/3}},$$

where $R = Bh/(B+2h)$ is the hydraulic radius and $n_M$ is Manning's friction coefficient.

The solution of Eq. (13) is approximated by means of a standard first-order Finite Volume scheme where a time splitting approach is adopted to separately treat the advective and the friction part of the mathematical model (Toro 2001). First, the vector $\mathbf{u}_i^n = \begin{pmatrix} h_i^n & h_i^n u_i^n \end{pmatrix}^T$ of the conserved variables in the cell $C_i$ at the time level $n$ is adjourned with the advective step (LeVeque 2002)

$$(16) \quad \mathbf{u}_i^* = \mathbf{u}_i^n - \frac{\Delta t}{\Delta x}\left[\mathbf{f}_{i+1/2}^- - \mathbf{f}_{i-1/2}^+\right],$$

then the implicit friction step

$$(17) \quad \mathbf{u}_i^{n+1} = \mathbf{u}_i^* - \Delta t \mathbf{S}_f\left(\mathbf{u}_i^{n+1}\right),$$

is used to calculate the vector $\mathbf{u}_i^{n+1}$ of the conserved variables at the time level $n + 1$ (Cozzolino et al. 2012). In Eqs. (16) and (17), $\Delta t$ is the time step, $\Delta x$ is the length of the cell, $\mathbf{f}_{i+1/2}^-$ is the contribution to the cell $C_i$ across the interface $i+1/2$ between $C_i$ and $C_{i+1}$, while $\mathbf{f}_{i+1/2}^+$ is the contribution to the cell $C_{i+1}$.

It is assumed that gates are located at cell interfaces, and a distinction is made between ordinary and gate interfaces. When the gate is absent, the interface is ordinary and the two contributes $\mathbf{f}_{i+1/2}^-$ and $\mathbf{f}_{i+1/2}^+$, which coincide, are approximated by means of the HLL numerical flux $\mathbf{f}_{i+1/2} = \mathbf{g}\left(\mathbf{u}_i^n, \mathbf{u}_i^n\right)$ described in Fraccarollo and Toro (1995). A limit depth $\varepsilon = 10^{-20}$ m is used to define dry cells. In dry cells the velocity is null and the momentum is not adjourned, while the numerical fluxes $\mathbf{f}_{i+1/2}^-$ and $\mathbf{f}_{i+1/2}^+$ between dry cells are null.

It remains to specify the treatment of $\mathbf{f}_{i+1/2}^-$ and $\mathbf{f}_{i+1/2}^+$ at gate interfaces, where the device is located. Two different procedures, called equilibrium and non-equilibrium approach, respectively, are described in the following.

**4.1 Classic equilibrium approach**

The name "equilibrium approach" refers to the fact that the numerical fluxes $\mathbf{f}^{-}_{i+1/2}$ and $\mathbf{f}^{+}_{i+1/2}$ at the gate are computed by imposing the gate equations in their original equilibrium form of Section 2 (steady state conditions). This approach seems well justified, since the states $\mathbf{u}_1$ and $\mathbf{u}_2$ immediately upstream and downstream of the gate, respectively, are constant for $t > 0$ (local steady state conditions) in the exact Riemann solution.

### *4.1.1 Algorithm structure*

We assume, without loss of generality, that the flow depth to the left of the gate is greater than the flow depth to the right, i.e., $h_i^n \geq h_{i+1}^n$ (the case with $h_i^n < h_{i+1}^n$ is easily managed after mirroring the reference framework). The following algorithm is inspired to that by Cozzolino et al. (2015), but obvious changes are made to consider the equations of Section 2:

*f*1) If $h_i^n < a$, the flow does not touch the gate lip (non-orifice flow) and $\mathbf{f}^{\pm}_{i+1/2} = \mathbf{f}_{i+1/2}$ is taken.

*f*2) If $h_i^n \geq a$, orifice flow regime is established. We assume that $h_u = h_i^n$ and calculate $q_F$, $C_c$, and $h_c^{\#}$, with Eqs. (2)-(4). Two different conditions are now possible:

    *f*2.1) if $h_{i+1}^n < h_c^{\#}$, free flow conditions are established; in this case

(18) $\mathbf{f}^{-}_{i+1/2} = \left( q_F \quad 0.5g\left(h_i^n\right)^2 + q_F^2/h_i^n \right)^T, \quad \mathbf{f}^{+}_{i+1/2} = \left( q_F \quad 0.5gh_c^2 + q_F^2/h_c \right)^T;$

    *f*2.2) if $h_{i+1}^n \geq h_c^{\#}$, submerged flow conditions are established; in this case, $q_S$ is calculated with Eq. (5) where $h_t = h_{i+1}^n$, and

(19) $\mathbf{f}^{-}_{i+1/2} = \left( q_S \quad 0.5g\left(h_i^n\right)^2 + q_S^2/h_i^n \right)^T, \quad \mathbf{f}^{+}_{i+1/2} = \left( q_S \quad 0.5g\left(h_{i+1}^n\right)^2 + q_S^2/h_{i+1}^n \right)^T.$

The computation of $C_c$ with the second of Eq. (3) is accomplished after that the parameter $\theta$ corresponding to the relative opening $a/h_u$ is found. Due to non-linearity of Eq. (3), this search is carried out iteratively using the bisection algorithm with initial guess $\theta_0 = -2.137(1-a/h_i^n)^2 + 43302(1-a/h_i^n) + 0.1897$. Notice that the Newton-Raphson algorithm is discarded due to failure for $a/h_i^n \to 1$.

### 4.1.2 Dam-break on dry bed

To test the capability of the equilibrium approach, the solution of the dam-break problem on dry bed (Test E1 of Table 1) is approximated with the numerical scheme of Eqs. (16)-(17) using computation parameters $\Delta x = 0.1$ m, $\Delta t = 0.002$ s, and $n_M = 0$ m$^{1/3}$/s. We recall that this problem exhibits three distinct exact solutions, but only the solution that lies on the L-M curve without gap is relevant (see Section 3.3). The numerical flow depth at time $t = 5$ is compared with the exact solution in Figure 13, where the computational results are represented with dots (only one in five dots is represented to improve clarity of the figure). The inspection shows that the equilibrium approach approximates the non-orifice solution and does not capture the relevant exact solution, which is characterized by orifice free flow with $\mathbf{u}_1 = \mathbf{u}_{F,h}$.

In the numerical scheme, the upstream flow depth $h_u$ coincides with the flow depth $h_1$ of the state $\mathbf{u}_1$ immediately to the left of the gate. The numerical results are further scrutinized in Figure 14, where the time-graphs of numerical unit-width discharge $q_F$ (Figure 14a), he numerical upstream flow depth $h_u$ (Figure 14b), and numerical relative opening $a/h_u$ (Figure 14c), are plotted. To interpret the figure, one must recall from Section 2.2 that $q_F$ is an increasing function of $h_u$ for $a/h_u < (a/h_u)_{\text{lim}} = 0.86$, while it is a decreasing function of $h_u$ for $a/h_u > (a/h_u)_{\text{lim}}$. The inspection of Figure 14a shows that $q_F$ rapidly increases from 0 to 1.05 m$^2$/s at the beginning of the transient, when the upstream flow depth is maximum, then it starts to decrease because $h_u$ decreases due to the channel emptying (Figure 14b). Interestingly, $a/h_u$ increases rapidly and crosses the threshold $(a/h_u)_{\text{lim}}$ at time $t = 0.08$ s (Figure

14c). This causes the discharge $q_F$ to stop its descent and start to increase again, accelerating the channel emptying and causing the final detachment of the flow from the gate lip at $t = 0.69$ s, after numerous oscillations caused by the alternating passage from orifice to non-orifice flow regime and *vice versa*. From the preceding analysis, it seems that the equilibrium approach overestimates $q_F$ during the initial phase of the dam-break transient, causing a too rapid decrease of $h_u$.

To shed light on this issue, consider the cell $C_i$ immediately upstream of the gate in the Finite Volume scheme of Eqs. (16)-(17). At the time level $n = 0$, the flow depth in $C_i$ coincides with the initial conditions flow depth ($h_i^0 = h_L$) while the unit-width discharge is null ($q_i^0 = 0$), implying that the total head in the cell $C_i$ is $H_i^0 = h_i^0$. During the first time-step, the mass-conservation component of Eq. (16) can be written in $C_i$ as

$$(20) \quad h_i^* = h_i^0 - \frac{\Delta t}{\Delta x}\left(q_{i+1/2} - q_{i-1/2}\right) = h_i^0 - \frac{\Delta t}{\Delta x} q_F$$

because the mass-flux $q_{i-1/2}$ between the cells $C_i$ and $C_{i-1}$ at time $t = 0$ s is null while the mass flux $q_{i+1/2}$ between the cells $C_i$ and $C_{i+1}$ coincides with $q_F$ The use of Eq. (2) for the evaluation of $q_F$ at time $t = 0$ s implicitly lies on the assumption that the total head $H_{i,req}^0 = h_i^0\left[1 + \left(C_c a/h_i^0\right)^2 / \left(1 + C_c a/h_i^0\right)\right] > h_i^0$ is available in the cell $C_i$ (see Section 2), but this assumption is not verified because $H_i^0 = h_i^0$. This explains why the discharge $q_F$ under the gate is overestimated at the beginning of the transient. A similar phenomenon will occur during the subsequent time steps because the head in the cell $C_i$ immediately upstream of the gate will generally differ from the required head associated to the discharge $q_F$ computed with Eq. (2).

The observations above suggest that the gate equation should be modified to take into account strong transients. A heuristic approach able to cope with this issue will be considered in the next Section.

[Insert Figure 13 about here]

[Insert Figure 14 about here]

**4.2 Non-equilibrium approach**

If we assume that the total head is invariant through the gate and relax the assumption of discharge invariance, we obtain the free flow equation (see Appendix A)

$$(21) \quad h_u + \frac{u_u^2}{2g} = h_c + \frac{q_F^2}{2gh_c^2},$$

where $u_u$ is the upstream velocity. Solving with respect $q_F$, one obtains

$$(22) \quad q_F = C_c a \sqrt{2g\left(h_u + \frac{u_u^2}{2g} - C_c a\right)}.$$

Some algebra shows that Eq. (22) coincides with Eq. (2) when steady state conditions, characterised by $q_F = h_u u_u$, are attained.

The discharge $q_F$ in Eq. (2) depends on the upstream depth only, and this may lead to an overestimation (or underestimation) of the actual upstream energy content during transients. On the other hand, Eq. (22) improves the evaluation of the upstream energy but neglects the obvious physical condition of discharge invariance through the gate. In the following, a compromise that compensates the two types of error is obtained by averaging the two formulations, which leads to

$$(23) \quad q_F = C_c a \sqrt{2gh_u} \left[ \frac{1}{2\sqrt{1+\frac{C_c a}{h_u}}} + \frac{1}{2}\sqrt{1 + \frac{1}{2}\frac{u_u^2}{gh_u} - \frac{C_c a}{h_u}} \right].$$

Again, Eq. (23) coincides with Eq. (2) when the discharge invariance is attained during steady state conditions. Note that Eq. (23) should be regarded as a numerical relaxation approach with a physical justification, and not as a novel physics equation.

The steps that constitute the "non-equilibrium" numerical approach coincide with the steps of the equilibrium-approach described in Section 4.1.1, with the only difference that the Eq. (23) with $h_u = h_i^n$ and $u_u = q_i^n/h_i^n$ is used instead of Eq. (2) to calculate the numerical discharge $q_F$. Congruently, the $q_F$ of Eq. (23) is also used to compute the limit tailwater depth $h_c^\#$ of Eq. (4) and the submerged flow discharge $q_S$ of Eq. (5).

*4.2.1 Numerical tests with exact solution*

The solution of the dam-break problem on dry bed (Test E1 of Table 1) is approximated with the non-equilibrium approach, using the same numerical parameters of Section 4.1.2. The corresponding flow depth at time $t = 5$ is compared with the exact solution in Figure 15, where the computational results are represented with dots (only one in five dots is represented). The inspection of the figure shows that the non-equilibrium approach approximates the relevant free flow solution. In particular, the flow depth jump through the gate is nicely captured, together with the strength and celerity of the moving waves. The inspection of Figure 16a shows that the non-equilibrium approach reaches the goal of reducing the overestimation of $q_F$ during the initial part of the transient. This allows to limit the decrease of $h_u$ (Figure 16b) and the increase of $a/h_u$ (Figure 16c), ensuring that the orifice flow regime is kept during the entire simulation.

[Insert Figure 15 about here]

[Insert Figure 16 about here]

The remaining tests of Table 1 are tackled with the non-equilibrium approach and computation parameters of Section 4.1.2, and the corresponding numerical results (flow depth at time $t = 5$ s) are compared with the exact solutions in Figure 17 (tests from E2 to E4), Figure 18 (tests E5 and E6), and Figure 19 (tests from E7 to E9). The inspection of the figures shows that the non-equilibrium numerical approach nicely approximates the exact solution, independent on the initial relative opening $a/h_L$.

[Insert Figure 17 about here]

[Insert Figure 18 about here]

[Insert Figure 19 about here]

### *4.2.2 Laboratory dam-break tests*

In the present Section, the numerical scheme of Eqs. (16)-(17), equipped with the "non-equilibrium" numerical approach for the gate discharge evaluation, is used to reproduce the results of the laboratory dam-break tests on dry bed carried out in a horizontal rectangular flume with plexiglass walls at the ICEA Department of the University of Padua (Lazzarin et al. 2023). The flume used during the laboratory experiments was $L = 6.0$ m long and $B = 0.30$ m wide, while a vertical sharp-crested plexiglass diaphragm, located at the centre of the channel, was used to simulate the presence of a sluice gate with fixed opening $a = 0.096$ m. A digital camera with recording rate $f = 24$ fps was used to record the experiments for subsequent image processing.

The six dam-break experiments carried out were characterised by different initial flow depths $h_L$, as reported in the second column of Table 2, while the corresponding values of $a/h_L$ are reported in the third column. The results of the experiments are resumed in the fourth and fifth column of the same table. In experiments from L1 to L3, the flow detached from the gate lip in a time interval minor than the camera recording time frame $t_f = 0.042$ s, with immediate establishment of a non-orifice flow regime. In experiments L4 and L5, orifice free flow conditions with upstream flow depth $h_u$ were present for a short time until detachment was completed at times $t^* = 0.5$ s and $t^* = 2$ s, respectively (see Table 2). Finally, Experiment L6 was characterized by stable orifice flow conditions.

The laboratory experiments are simulated with computation parameters $\Delta x = 0.01$ m, $\Delta t = 0.001$ s, and $n_M = 0.01$ m$^{1/3}$/s, imposing wall boundary conditions to the left end of the flume and a free fall to the right. The corresponding results are summarized in the sixth and seventh column of Table 2, while the time histories of the depth $h_u$ for all the tests are plotted in Figure 20. The inspection of Table 2 and Figure 20 shows that the detachment of the flow from the gate lip in numerical experiments L1 to L3, like the corresponding laboratory experiments, is completed in a time $t^* < t_f$, leading to immediate non-orifice flow regime. Corresponding to the laboratory experiment results, orifice free flow of numerical tests L4 and L5 is kept for a short time until non-orifice flow conditions are established. Finally, stable free flow conditions are simulated during the numerical experiment L6. In all the cases where orifice flow regime is established, the flow depth $h_u$ immediately upstream of the gate is comparable to the corresponding laboratory experimental depth. In conclusion, the numerical simulations show that the model, equipped with a non-equilibrium approach for the computation of the discharge under the gate and a simple friction model, can reproduce the essentials of the laboratory experiments and capture the limit between orifice and non-orifice regimes.

For the sake of comparison, the dam-break exact solutions obtained with the methods of Section 3.1 and the corresponding Finite Volume solution without friction are reported in Table 3. The inspection of Table 3 confirms that the numerical model without friction is able to nicely capture the corresponding exact solutions, as already deduced in Section 4.2.1. More interestingly, the comparison with Table 2 shows that, contrarily to the numerical results with friction, the frictionless exact and numerical solutions do not reproduce all the experimental flow regimes. It can be deduced that the friction has a decisive influence in determining the numerical simulation results in the case of rapid transients with partially lifted sluice gates. The discussion of the friction influence on the laboratory dam-break solutions will be tackled in Section 6.

[Insert Table 2 about here]

[Insert Figure 20 about here]

[Insert Table 3 about here]

**5. Two-dimensional framework**

The 2-d SWE model with uneven bed elevation and friction can be written as (Audusse and Bristeau 2005)

$$(24) \quad \frac{\partial \mathbf{U}}{\partial t} + \frac{\partial \mathbf{F}(\mathbf{U})}{\partial x} + \frac{\partial \mathbf{G}(\mathbf{U})}{\partial y} = \mathbf{S}_0(\mathbf{U}) - \mathbf{S}_f(\mathbf{U}).$$

where the conserved variable vector $\mathbf{U}$, and the flux vectors $\mathbf{F}$ and $\mathbf{G}$ along $x$ and $y$, respectively, are defined as

$$(25) \quad \begin{aligned} \mathbf{U} &= \begin{pmatrix} h & hu & hv \end{pmatrix}^T \\ \mathbf{F}(\mathbf{U}) &= \begin{pmatrix} hu & 0.5gh^2 + hu^2 & huv \end{pmatrix}^T, \\ \mathbf{G}(\mathbf{U}) &= \begin{pmatrix} hv & huv & 0.5gh^2 + hv^2 \end{pmatrix}^T \end{aligned}$$

while the vectors of the source terms $\mathbf{S}_0$ and $\mathbf{S}_f$ are defined as

$$(26) \quad \mathbf{S}_0(\mathbf{U}) = \begin{pmatrix} 0 & -gh\frac{\partial z_b}{\partial x} & -gh\frac{\partial z_b}{\partial y} \end{pmatrix}^T, \quad \mathbf{S}_f(\mathbf{U}) = \begin{pmatrix} 0 & ghS_{f,x} & ghS_{f,y} \end{pmatrix}^T.$$

In Eqs. (24)-(26), the meaning of the symbols is as follows: $h$ is the flow depth; $u$ and $v$ are the components of the velocity along $x$ and $y$, respectively; $z_b$ is the bed elevation; $S_{f,x}$ and $S_{f,y}$ are the components of the friction slope along $x$ and $y$, respectively. In practical applications, $S_{f,x}$ and $S_{f,y}$ are computed with the Manning's formula:

$$(27) \quad S_{f,x} = \frac{n_M^2 u \sqrt{u^2 + v^2}}{h^{4/3}}, \quad S_{f,y} = \frac{n_M^2 v \sqrt{u^2 + v^2}}{h^{4/3}}.$$

In the following, it will be shown how the 1-d gate model of Section 2 can be adjusted for implementation in 2-d SWE models, and a 2-d example application will be presented.

**5.1 Gate model for 2-d flows**

Consider the plan-view of a sluice gate as represented in Figure 21, where the gate is aligned with the $y$-axis of the fixed reference $Oxy$ (Figure 21a). The flow particle approaching the gate has velocity components $u_u$ and $v_u$ along $x$ and $y$, respectively. After the passage under the gate, the components of the velocity become $u_d$ and $v_d$, respectively. Mass conservation at the gate implies that $q_g = h_u u_u = h_d u_d$, where $q_g$ is the unit-width discharge under the gate, while $h_u$ and $h_d$ are the flow depths upstream and downstream, respectively. If the gate is frictionless, it exerts no action on the flow particle along the $y$-axis, implying that the flow particle has no acceleration along $y$ while passing under the gate. This supplies $v_u = v_d$ (invariance of the transverse velocity, Figure 21a) and $q_g v_u = q_g v_d$ (invariance of the transverse momentum flux).

Consider now a moving reference $O'xy'$ that translates along $y$ with uniform velocity $v_u$ and such that $y' = y - v_u t$. In this moving reference, the particle passes perpendicularly under the gate and has no velocity component along $y'$ (Figure 21b), implying that the gate equations in the moving reference $O'xy'$ coincide with the 1-d gate equations (Section 2). Of course, the shift of $O'xy'$ along $y$ leaves $q_g$ unchanged because the $x$-components $u_u$ and $u_d$ of the velocity are unaffected. In addition, we observe that the passage from the *Oxy* reference to the *O'xy'* reference does not introduce inertial forces because *O'xy'* moves with uniform velocity with respect to *Oxy*. From the preceding, it can be concluded what follows:

- in the 2-d case, the unit-width discharge under the gate can be calculated using the 1-d gate equations of Section 2;
- when the non-equilibrium numerical approach is applied for the numerical computation of the unit-width discharge under the gate, the velocity appearing in Eq. (23) coincides with the component of the upstream velocity that is normal to the gate;
- the forces that the gate exerts on upstream and downstream flows coincide with the forces calculated in the 1-d case, and the transverse momentum flux under the gate reduces to $q_g v_u$.

[Insert Figure 21 about here]

**5.2 Two-dimensional numerical modelling**

The solution of the 2-d SWE of Eq. (24) is approximated by means of a first-order Finite Volume scheme on unstructured triangular grid, where a time splitting approach is adopted to separately treat the advective and the friction part of the mathematical model (Toro 2001). In the cell $C_i$, the vector $\mathbf{U}_i^n = \begin{pmatrix} h_i^n & h_i^n u_i^n & h_i^n v_i^n \end{pmatrix}^T$ of the conserved variables at the time level $n$ is first adjourned with the explicit advective step

$$(28) \quad \mathbf{U}_i^* = \mathbf{U}_i^n - \frac{\Delta t}{|C_i|} \sum_{j \in K(i)} |l_{ij}| \mathbf{R}_{ij}^{-1} \left[ \delta_{ij} \mathbf{F}_{ij}^- \left( \mathbf{R}_{ij} \hat{\mathbf{U}}_{ij}^n, \mathbf{R}_{ij} \hat{\mathbf{U}}_{ji}^n \right) + (1-\delta_{ij}) \mathbf{G}_{ij}^- \left( \mathbf{R}_{ij} \mathbf{U}_i^n, \mathbf{R}_{ij} \mathbf{U}_j^n \right) \right],$$

while the implicit friction step

(29) $\mathbf{U}_i^{n+1} = \mathbf{U}_i^* - \Delta t \mathbf{S}_f \left( \mathbf{U}_i^{n+1} \right),$

is subsequently used to calculate the vector $\mathbf{U}_i^{n+1}$ of the conserved variables at the time level $n + 1$.

In Eqs. (28)-(29), the meaning of the symbols is as follows: $\mathbf{U}_i^*$ is the adjourned vector of the conserved variables in the cell $C_i$ after the advective step; $|C_i|$ is the area of the cell $C_i$; $K(i)$ is the set of the cells that are contiguous to $C_i$; $|l_{ij}|$ is the length of the interface $l_{ij}$ between the cells $C_i$ and $C_j$, where $\mathbf{n}_{ij} = \left( n_{ij,x} \quad n_{ij,y} \right)^T$ is the unit-length vector normal to $l_{ij}$ and directed from $C_i$ to $C_j$; $\delta_{ij}$ is a binary indicator that is equal to 1 if a gate is not present on the interface $l_{ij}$ (ordinary interface), while it is equal to 0 if the gate is located on $l_{ij}$ (gate interface); $\mathbf{F}_{ij}^-$ is the flux and bed slope contribution, projected along $\mathbf{n}_{ij}$, of the ordinary interface $l_{ij}$ to $C_i$, while $\mathbf{G}_{ij}^-$ is the contribution of the gate interface; finally, $\mathbf{R}_{ij}$ is a rotation matrix defined as (Toro 2001)

(30) $\mathbf{R}_{ij} = \begin{pmatrix} 1 & 0 & 0 \\ 0 & n_{ij,x} & n_{ij,y} \\ 0 & -n_{ij,y} & n_{ij,x} \end{pmatrix}.$

The matrix $\mathbf{R}_{ij}$ allows the passage from the global reference framework $Oxy$ to a local reference whose axes are aligned with the interface between $C_i$ and $C_j$. In Eq. (28), the vectors $\hat{\mathbf{U}}_{ij}^n = \left( \hat{h}_{ij}^n \quad \hat{h}_{ij}^n u_i^n \quad \hat{h}_{ij}^n v_i^n \right)^T$ and $\hat{\mathbf{U}}_{ji}^n = \left( \hat{h}_{ji}^n \quad \hat{h}_{ji}^n u_j^n \quad \hat{h}_{ji}^n v_j^n \right)^T$ are obtained from the vectors $\mathbf{U}_i^n$ and $\mathbf{U}_j^n$, respectively, by applying the hydrostatic reconstruction for the treatment of bed elevation terms (Audusse et al. 2004). The corresponding flow depths are defined as $\hat{h}_{ij}^n = \left( h_i^n + z_{b,i} - z_{b,j} \right)_+$ and $\hat{h}_{ji}^n = \left( h_j^n + z_{b,j} - z_{b,i} \right)_+$, respectively, where $z_{b,i}$ and $z_{b,j}$ are the cell-averaged bed elevations in $C_i$ and $C_j$.

*5.2.1 Computation of ordinary interface contributions*

At ordinary interfaces, where the gate is not present, a simplified HLLC approximate Riemann solver (Toro 2001) is used to solve the local SWE plane Riemann problem while the hydrostatic reconstruction approach by Audusse et al. (2004) is adopted to cope with the source term $\mathbf{S}_0(\mathbf{U})$. For this reason, it is possible to write

$$(31) \quad \mathbf{F}_{ij}^{-}\left(\mathbf{R}_{ij}\hat{\mathbf{U}}_{i}^{n}, \mathbf{R}_{ij}\hat{\mathbf{U}}_{j}^{n}\right) = \mathbf{F}^{HLLC}\left(\mathbf{R}_{ij}\hat{\mathbf{U}}_{i}^{n}, \mathbf{R}_{ij}\hat{\mathbf{U}}_{j}^{n}\right) + \begin{pmatrix} 0 \\ \dfrac{g}{2}\left[\left(h_{i}^{n}\right)^{2} - \left(\hat{h}_{ij}^{n}\right)^{2}\right]\mathbf{n}_{ij} \end{pmatrix},$$

where the HLLC numerical flux $\mathbf{F}^{HLLC}$ corresponding to the SWE plane Riemann problem is calculated using the projections $\mathbf{R}_{ij}\hat{\mathbf{U}}_{ij}^{n}$ and $\mathbf{R}_{ij}\hat{\mathbf{U}}_{ji}^{n}$ of the reconstructed conserved variables $\hat{\mathbf{U}}_{ij}^{n}$ and $\hat{\mathbf{U}}_{ji}^{n}$, respectively.

*5.2.2 Computation of gate interface contributions*

For the sake of simplicity, it is assumed that the bed is horizontal under gates and that the opening is $a$. When $\max\left(h_{i}^{n}, h_{j}^{n}\right) < a$, non-orifice flow conditions are established and $\mathbf{G}_{ij}^{-}$ reduces to

$$(32) \quad \mathbf{G}_{ij}^{-}\left(\mathbf{R}_{ij}\mathbf{U}_{i}^{n}, \mathbf{R}_{ij}\mathbf{U}_{j}^{n}\right) = \mathbf{F}^{HLLC}\left(\mathbf{R}_{ij}\mathbf{U}_{i}^{n}, \mathbf{R}_{ij}\mathbf{U}_{j}^{n}\right).$$

In Section 5.1, we have shown that a locally 1-d approach can be used to compute the discharge $q_g$ under the gate in the case of 2-d flows. If orifice flow conditions are established because $\max\left(h_{i}^{n}, h_{j}^{n}\right) \geq a$, two different conditions are possible. If $h_{i}^{n} \geq h_{j}^{n}$, the flow is from the cell $C_i$ (upstream) to the cell $C_j$ (downstream), and one has

$$(33)\ \mathbf{G}_{ij}^{-}\left(\mathbf{R}_{ij}\mathbf{U}_{i}^{n},\mathbf{R}_{ij}\mathbf{U}_{j}^{n}\right)=\left(q_{g}\quad 0.5g\left(h_{i}^{n}\right)^{2}+q_{g}^{2}/h_{i}^{n}\quad q_{g}v_{u}\right)^{T},$$

where $q_g = q_F$ if orifice free flow is established ($h_j^n < h_c^{\#}$), otherwise $q_g = q_S$ in the case of submerged flow ($h_j^n \geq h_c^{\#}$). The free flow discharge $q_F$ is computed with the non-equilibrium approach of Eq. (23), where $h_u = h_i^n$ and $u_u = n_{ij,x}u_i^n + n_{ij,y}v_i^n$ (normal component of the velocity). Congruently, the $q_F$ of Eq. (23) is also used to compute the limit tailwater depth $h_c^{\#}$ of Eq. (4) and the submerged flow discharge $q_S$ of Eq. (5), where $h_t = h_j^n$. The transverse component of the upstream velocity is calculated using $v_u = -n_{ij,y}u_i^n + n_{ij,x}v_i^n$.

If $h_i^n < h_j^n$, the flow is directed from the cell $C_j$ (upstream) to the cell $C_i$ (downstream), and

$$(34)\ \mathbf{G}_{ij}^{-}\left(\mathbf{R}_{ij}\mathbf{U}_{i}^{n},\mathbf{R}_{ij}\mathbf{U}_{j}^{n}\right)=-\left(q_{F}\quad 0.5gh_{c}^{2}+q_{F}^{2}/h_{c}\quad q_{F}v_{u}\right)^{T},$$

is used in the case of free flow ($h_i^n < h_c^{\#}$), while

$$(35)\ \mathbf{G}_{ij}^{-}\left(\mathbf{R}_{ij}\mathbf{U}_{i}^{n},\mathbf{R}_{ij}\mathbf{U}_{j}^{n}\right)=-\left(q_{S}\quad 0.5g\left(h_{i}^{n}\right)^{2}+q_{S}^{2}/h_{i}^{n}\quad q_{S}v_{u}\right)^{T}$$

is used in the case of submerged flow ($h_i^n \geq h_c^{\#}$). In Eq. (34), the free flow discharge $q_F$ is computed with Eq. (23), where $h_u = h_j^n$ and $u_u = n_{ij,x}u_j^n + n_{ij,y}v_j^n$, while the submerged flow discharge $q_S$ is computed with Eq. (5) where $h_t = h_i^n$. The transverse component of the upstream velocity is calculated using $v_u = -n_{ij,y}u_j^n + n_{ij,x}v_j^n$.

*5.2.3 Two-dimensional idealized detention basin*

The 2-d numerical model described above is applied to simulate the filling and emptying of a detention basin (see Figure 22). The detention basin consists of a rectangular reservoir with length $L$ = 40 m and width $W$ = 42 m (Figure 22a). The inflow and the outflow consist of two rectangular channels with width $B$ = 2 m, which are connected to a longitudinal trapezoidal channel with base width $B_{base}$ = 2 m, height $H$ = 1 m, and top width $B_{top}$ = 4 m (Figure 22c). The outflow channel has null slope, while the approaching channel has slope $S_{0,i}$ = 0.005 and the trapezoidal channel has slope $S_{0,t}$ = 0.001 (Figure 22b). A rectangular sluice gate (Figure 22c), whose width is $B_{base}$ = 2 m and opening $a$ = 0.55 m, is present at the end of the trapezoidal channel. Free fall boundary conditions are imposed at the end of the outlet channel, while inflow boundary conditions are imposed at the inlet channel. A uniform Manning's friction coefficient $n_M$ = 0.013 s/m$^{1/3}$ is used in the entire physical domain.

The 2-d physical domain is discretized with a triangular unstructured grid, with sides $s$ = 0.25 m long in the gate region, while $s$ = 0.80 m is used at the basin walls. The initial conditions correspond to steady flow with discharge $Q_{in}$ = 1.56 m$^3$/s and inlet flow depth $h_{in}$ = 0.5 m. From $t$ = 0 s, the inflow discharge and flow depth are varied following the hydrographs of Figure 23, where the inflow discharge and the corresponding flow depth are represented with a thin and a thick black line, respectively.

The free-surface profile along the longitudinal axis corresponding to the initial condition is represented in Figure 24a (flow from left to right). From the figure, it can be observed that the inflow supercritical flow is reversed into subcritical by a hydraulic jump located in the inlet channel. An additional increase of the flow depth is found at the passage from the inlet channel to the detention basin, while the free-surface decreases at the passage from the detention basin to the outlet channel. At this point, the free surface does not touch the gate lip and non-orifice flow conditions are established.

For $t > 0$ s, the detention basin starts to fill (Figure 24b, $t = 300$ s) until the maximum outflow discharge is attained at $t = 1320$ s (Figure 24c), after which it slowly empties (Figure 24d, $t = 1800$ s). In Figure 25, the outflow hydrograph (with peak discharge $Q_{p,o} = 4.27$ m³/s) is compared with the inflow hydrograph (peak discharge $Q_{p,i} = 8.04$ m³/s). The inspection of the figure shows that the detention basin attains a lamination efficiency $\eta = 1 - Q_{p,o}/Q_{p,i} = 0.47$ for the inflow hydrograph considered.

For the sake of comparison, the exercise is repeated without gate. The inspection of Figure 25, where the outflow discharge for the case without gate is represented with a dashed line, highlights the dramatic increase of efficiency introduced by the device. In fact, the peak outflow without gate is $Q_{p,o} = 6.79$ m³/s, corresponding to lamination efficiency $\eta = 0.16$.

[Insert Figure 22 about here]

[Insert Figure 23 about here]

[Insert Figure 24 about here]

[Insert Figure 25 about here]

## 6. Discussion

The solution disambiguation criterion proposed in Section 3.3 is based on the concept of existence and uniqueness of frictionless dam-break exact solutions for general initial conditions. On the other hand, the disambiguation criterion proposed by Lazzarin et al. (2023), which has been validated by means of the laboratory dam-break experiments of Table 2, is based on the stability of the sluice gate equations of Section 2 only, without regard for the existence and the uniqueness of the dam-break solutions for general initial conditions. The last criterion requires that the ratio $a/h_u$ between the gate opening $a$ and the flow depth $h_u$ immediately upstream of the gate satisfies the condition $a/h_u < (a/h_u)_{\lim}$, where $(a/h_u)_{\lim} = 0.86$.

To compare the two disambiguation criteria, the flow depth $h_1$ corresponding to the state $\mathbf{u}_1$ in the exact solution to the dam-break problem on dry bed for different values of $a$ and $h_L$ is considered. In Figure 26a, which is obtained from Figure 4 after the application of the disambiguation criterion of Section 3.3, the ratio $h_1/h_L$ is represented as a function of $a/h_L$. The inspection of the figure shows that the dam-break on dry bed supplies non-orifice flow regime when $a/h_L > 0.495$, while orifice flow regime is obtained for $a/h_L \in \left]0, 0.495\right]$ (see Section 3.3). In Figure 26b, a similar diagram is plotted after the application of the disambiguation criterion by Lazzarin et al. (2023). In the case of the dam-break problem, this criterion is equivalent to the condition $a/h_1 < \left(a/h_u\right)_{\lim}$, which implies that non-orifice flow regime is obtained for $a/h_L > 0.491$, while orifice flow regime is obtained for $a/h_L \in \left]0, 0.491\right]$. The comparison between Figures 26a and 26b shows that the two criteria, while based on very different assumptions, lead to results that differ only in the very narrow region $a/h_L \in \left[0.491, 0.495\right]$, where the laboratory experiment L5 of Table 2 falls. For this experiment, the criterion by Lazzarin et al. (2023) predicts non-orifice flow regime, which is confirmed by the laboratory results, while the disambiguation criterion of Section 3.3 predicts orifice flow regime (see Table 3, fifth column).

Although the last observation is apparently negative for its credibility, we notice that the disambiguation criterion of Section 3.3 is expressly formulated for exact and numerical solutions without friction. For this reason, its direct application is inappropriate in real world cases while it comes useful in the construction of numerical models based on the local solution of a Riemann problem. This is demonstrated by the 1-d numerical model of Section 4.2, which satisfies the disambiguation criterion of Section 3.3 for very idealistic cases without friction and nicely reproduces the L5 laboratory results (sixth column of Table 2) when the friction is added. To shed light on this apparent contradiction, in Figure 27 we contrast the time history of $h_u$ in the numerical simulations without and with friction for the experiment L5 of Table 2. In the numerical simulation without friction (Figure 27a), the flow depth $h_u$ rapidly drops until it starts to increase and tends to the dam-break exact solution. The numerical simulation with friction (Figure 27b) follows a similar trend up to $t = 0.5$ s, where $h_u$ attains the maximum $h_u = 0.108$ m, which corresponds to $a/h_u = 0.89 > (a/h_u)_{lim}$. After this point, $h_u$ starts to slowly decrease until a rapid drop detaches the flow from the gate lip, congruently with the laboratory experiment. The rapid detachment is easily explained by recalling that a decrease of $h_u$ causes the increase of the discharge under the gate when $a/h_u > (a/h_u)_{lim}$ (see Section 2), which in turn exacerbates the fall of $h_u$ like a snowball effect. It is evident that the loss of energy introduced by the friction is responsible for the $h_u$ decrease that undermines the orifice flow stability. The last observation is confirmed by the inspection of Figure 28, where the flow depths supplied at time $t = 1.7$ s by the numerical model of Section 4.2 are plotted for the L5 dam-break of Table 2. The comparison between Figure 28a (without friction) and Figure 28b (with friction) confirms that the influence of the friction on the dam-break solution is dramatic not only at the propagating wave toe, as commonly reported in the literature (Dressler 1952, Hogg and Pritchard 2004), but also at the gate position. In conclusion, a purely mathematical criterion for the disambiguation of multiple solutions to the gate Riemann problem like the one presented in this paper is not an obstacle to the simulation of realistic shallow water transients if additional effects like the friction are added in numerical computations.

The inspection of Figures 27 and 28 suggest a final consideration. The exact dam-break solutions of Section 3 are characterised by the existence of discernible uniform states, one of which is the state $\mathbf{u}_1$ immediately upstream of the gate. This idealisation, which is confirmed by the inviscid numerical solution of Figure 27a, should be assumed with some prudence. As evidenced by Figures 27a and 28a, the region immediately upstream of the gate is only approximately uniform in space and constant in time when the friction is present. While this is not a serious obstacle to the application of the Riemann problem solution in numerical schemes, the present observation suggests that the interpretation of fast transient laboratory experiments by means of the same framework should be carried out with great attention.

[Insert Figure 26 about here]

[Insert Figure 27 about here]

[Insert Figure 28 about here]

## 7. Conclusions

In the present paper, an improved solution of the dam-break problem at partially lifted sluice gates has been presented. This novel solution assumes not only the dependence of the gate contraction coefficient on the upstream flow depth (Defina and Susin 2003), but also recent developments for the definition of a physically congruent submerged flow equation (Bijankhan et al. 2012b). The improvement of the Riemann problem physical representation amends the limitations of the preceding work by Cozzolino et al. (2015), namely the lack of solution existence for certain values of the initial downstream flow depth.

As common for the Riemann problem of the Shallow water Equations at geometric discontinuities and hydraulic structures, there are initial conditions for which the solution is multiple, and a disambiguation criterion must be introduced to pick up a physically congruent choice among the alternatives. In the present work, a disambiguation criterion based on the continuous dependence of the solution on the initial conditions allows to single out a well-posed solution. Interestingly, this criterion supplies results that slightly differ from the ones obtained with the disambiguation criterion by Lazzarin et al. (2023), and the corresponding discrepancies are discussed.

Moreover, it is shown that the classic steady state gate model from the literature may lead to the overestimation of the discharge issuing under the gate during dam-break numerical computations. For this reason, a relaxed form of the gate equations, here called *non-equilibrium approach*, has been introduced and used in two novel Finite Volume schemes for the approximate solution of the Riemann problem at sluice gates. The 1-d Finite Volume numerical scheme with the non-equilibrium approach captures the exact solutions to the sluice gate dam-break problem introduced in the present work, picking up the relevant solution among the alternatives when multiple solutions are possible. The same numerical scheme with friction, reproduces with good accuracy the laboratory dam-break results by Lazzarin et al. (2023). It follows that the numerical scheme can distinguish between the dam-break initial conditions that either lead to orifice flow under the gate or to a flow that is detached from the gate lip in the flume experiments. Interestingly, the comparison between the numerical simulations with and without friction shows that the friction may have a role in the inception of the instability phenomena that lead to the detachment of the flow from the gate lip, and this has a consequence in the interpretation of laboratory experiments.

Finally, a 2-d Finite Volume scheme based on the non-equilibrium approach is used to simulate the filling and emptying of a detention basin with complicate topography and a sluice gate located at its downstream end, demonstrating how the novel findings can be promptly used in real-world applications.

**Acknowledgements**

The research project "Sistema di supporto decisionale per il progetto di casse di espansione in linea in piccoli bacini costieri" was funded by the Italian Ministry for Environment, Land and Sea Protection through the funding program "Metodologie per la valutazione dell'efficacia sulla laminazione delle piene in piccoli bacini costieri di sistemi di casse d'espansione in linea realizzate con briglie con bocca tarata".

**Appendix A. The non-equilibrium gate formula**

In the present Appendix we give a physical justification to Eq. (21). If we assume that the flow under the gate in free flow conditions is inviscid, the flow field satisfies the equation (Rouse 1946)

$$\text{(A.1)} \quad \frac{1}{g}\frac{\partial v_s}{\partial t} + \frac{\partial}{\partial s}\left(z + \frac{p}{\gamma} + \frac{v^2}{2g}\right) = 0$$

along the particle trajectory under the gate from the upstream position A to the position B at the *vena contracta* (Figure 29). In Eq. (A.1), $s$ is the local abscissa along the trajectory, $v$ is the particle velocity modulus, $v_s$ is the component of the velocity along the trajectory, $z$ is the elevation of the particle above the datum, $p$ is the local pressure, and $\gamma$ is the fluid specific weight. If we integrate in space from A to B and assume that the local acceleration is negligible, Eq. (A.1) reduces to

$$(A.2) \quad \left(z + \frac{p}{\gamma} + \frac{v^2}{2g}\right)_A = \left(z + \frac{p}{\gamma} + \frac{v^2}{2g}\right)_B.$$

If we further assume that the variability of the flow velocity in the cross-section is negligible and the flow is gradually varied in A and B, Eq. (A.2) can be rewritten as

$$(A.3) \quad h_u + \frac{u_u^2}{2g} = h_c + \frac{q_F^2}{2gh_c^2},$$

where $u_u$ is the upstream velocity. Eq. (A.3) coincides with Eq. (21).

[Insert Figure 29 about here]

**Tables List**9

Table 1. Initial conditions of the dam-break problems with exact solution.

Table 2. Laboratory dam-break tests: initial conditions, laboratory results, numerical solution with friction. An asterisk denotes the passage from orifice free flow to non-orifice flow regime.

Table 3. Laboratory dam-break tests: initial conditions, exact frictionless solution, numerical frictionless solution.

**Figures List**

Figure 1. Relationship between flow and sluice gate: orifice free-flow conditions (a); orifice submerged flow conditions (b); non-orifice flow regime (c).

Figure 2. Contraction coefficient $C_c$ (a) and squared Froude number $F_F^2$ (b) as functions of the relative opening $a/h_u$.

Figure 3. Dam-break on dry bed with $a = 0.47$ m and $h_L = 1$ m (Test E1 of Table 1), exact solutions (flow depth) at $t = 5$ s: non-orifice solution (a); free flow solution with $\mathbf{u}_1 = \mathbf{u}_{F,l}$ (b); free flow solution with $\mathbf{u}_1 = \mathbf{u}_{F,h}$ (c).

Figure 4. Dam-break on dry bed: relative flow depth $h_1/h_L$ as a function of the initial relative opening $a/h_L$.

Figure 5. Exact solutions (flow depth) at $t = 5$ s for the dam-break tests of Table 1 with $a/h_L = 0.2$: Test E2 (a); Test E3 (b); Test E4 (c).

Figure 6. Construction of the exact solution for the dam-break with $a/h_L = 0.2$: L-M curve (a); solution of Test E2 (b); solution of Test E3 (c); solution of Test E4 (d).

Figure 7. Exact solutions (flow depth) at $t = 5$ s for the dam-break tests of Table 1 with $a/h_L = 0.6$: Test E5 (a); Test E6 (b).

Figure 8. Construction of the exact solution for the dam-break with $a/h_L = 0.6$: L-M curve (a); solution of Test E5 (b); solution of Test E6 (c).

Figure 9. Exact solutions (flow depth) connected to $\mathbf{u}_{F,h}$ at $t = 5$ s for the dam-break tests of Table 1 with $a/h_L = 0.47$: Test E7 (a); Test E8 (b); Test E9 (c).

Figure 10. Construction of the exact solutions for the dam-break with $a/h_L = 0.47$: L-M curve connected to $\mathbf{u}_{F,h}$ (a); solution of Test E7 (b); solution of Test E8 (c); solution of Test E9 (d).

Figure 11. L-M curves with a gap for the dam-break with $a/h_L = 0.47$: L-M curve related to $\mathbf{u}_{F,l}$ (a); L-M curve related to the non-orifice solution of the dam-break on dry bed (b).

Figure 12. Unit-width discharges $q_{R,1}$ and $q_F$ for the construction of the TC curve in the case $a/h_L = 0.47$: L-M curve related to $\mathbf{u}_{F,h}$ (a); L-M curve related to $\mathbf{u}_{F,l}$ (b); L-M curve related to the non-orifice solution of the dam-break on-dry bed (c).

Figure 13. Dam-break on dry bed with $a = 0.47$ m and $h_L = 1$ m (Test E1 of Table 1). Flow depth at $t = 5$ s: exact solution (thin black line); numerical solution with the equilibrium approach (dots).

Figure 14. Dam-break on dry bed with $a = 0.47$ m and $h_L = 1$ m (Test E1 of Table 1) with the equilibrium approach. Time graphs of unit-width discharge under the gate (a); flow depth upstream of the gate (b); relative opening (c).

Figure 15. Dam-break on dry bed with $a = 0.47$ m and $h_L = 1$ m (Test E1 of Table 1). Flow depth at $t = 5$ s: exact solution (thin black line); numerical solution with the non-equilibrium approach (dots).

Figure 16. Dam-break on dry bed with $a = 0.47$ m and $h_L = 1$ m (Test E1 of Table 1) with the non-equilibrium approach. Time graphs of unit-width discharge under the gate (a); flow depth upstream of the gate (b); relative opening (c).

Figure 17. Comparison between exact (thin black line) and numerical solution with the non-equilibrium approach (dots) for the dam-break tests of Table 1 with $a/h_L = 0.2$. Flow depth at $t = 5$ s: Test E2 (a); Test E3 (b); Test E4 (c).

Figure 18. Comparison between exact (thin black line) and numerical solution with the non-equilibrium approach (dots, one in five is represented to enhance the clarity of the plot) for the dam-break tests of Table 1 with $a/h_L = 0.6$. Flow depth at $t = 5$ s: Test E5 (a); Test E6 (b).

Figure 19. Comparison between exact (thin black line) and numerical solution with the non-equilibrium approach (dots, one in five is represented to enhance the clarity of the plot) for the dam-break tests of Table 1 with $a/h_L = 0.47$. Flow depth at $t = 5$ s: Test E7 (a); Test E8 (b); Test E9 (c).

Figure 20. Numerical time histories (thin black line) of the upstream flow depth for the dam-break experiments of Table 2: Test L1 (a); Test L2 (b); Test L3 (c); Test L4 (d); Test L5 (e); Test L6 (f). The position of the gate lip is represented with a dashed line.

Figure 21. Sluice gate plan-view: fixed reference $Oxy$ (a); reference $O'xy'$ translating with uniform velocity $v_u$ (b).

Figure 22. Geometry of the detention basin: plan view (a); longitudinal section A-A (b); transverse cross-section B-B (c). Distorted representation.

Figure 23. Inflow to the detention basin: discharge (thin black line) and flow depth (thick black line) hydrographs.

Figure 24. Free-surface profile along the detention basin longitudinal axis at different times: $t = 0$ s (a); $t = 300$ s (b); $t = 1320$ s (c); $t = 1800$ s.

Figure 25. Detention basin inflow and outflow discharges: inflow discharge (thin black line); outflow discharge with gate (thick black line); outflow discharge without gate (dashed line). The arrows individuate the instants corresponding to times $t = 0$ s (a), $t = 300$ s (b), $t = 1320$ s (c), $t = 1800$ s (d), respectively.

Figure 26. Admissible solutions of the dam-break on dry bed with partially lifted sluice gate: disambiguation criterion of Section 3.3 (a); disambiguation criterion by Lazzarin et al. (2023).

Figure 27. Numerical solution of the L5 dam-break problem (Table 2) with the numerical model of Section 4.2 (thin black line). Flow depth immediately upstream the gate: simulation without friction (a); simulation with friction (b). The position of the gate lip is represented with a dashed line.

Figure 28. Numerical solution of the L5 dam-break problem (Table 2) with the numerical model of Section 4.2. Flow depth along the flume at time $t = 1.7$ s: simulation without friction (a); simulation with friction (b).

Figure 29. Physical justification of the non-equilibrium approach for the computation of the discharge issuing under the gate: particle trajectory under the gate from the upstream position A to the position B at the *vena contracta*.

Table 1. Initial conditions of the dam-break problems with exact solution.

| Test | $h_L$ (m) | $h_R$ (m) | $a$ (m) | $a/h_L$ | Notes |
|---|---|---|---|---|---|
| E1 | 1.0 | 0 | 0.47 | 0.47 | Figure 3 |
| E2 | 1.0 | 0.002 | 0.2 | 0.2 | Figures 5a, 6b |
| E3 | 1.0 | 0.2 | 0.2 | 0.2 | Figures 5b, 6c |
| E4 | 1.0 | 0.6 | 0.2 | 0.2 | Figures 5c, 6d |
| E5 | 1.0 | 0.25 | 0.6 | 0.6 | Figures 7a, 8b |
| E6 | 1.0 | 0.6 | 0.6 | 0.6 | Figures 7b, 8c |
| E7 | 1.0 | 0.002 | 0.47 | 0.47 | Figures 9a, 10b |
| E8 | 1.0 | 0.2 | 0.47 | 0.47 | Figures 9b, 10c |
| E9 | 1.0 | 0.6 | 0.47 | 0.47 | Figures 9c, 10d |

Table 2. Laboratory dam-break tests: initial conditions, laboratory results, numerical solution with friction. An asterisk denotes the passage from orifice free flow to non-orifice flow regime.

| Test | $h_L$ (m) | $a/h_L$ | Laboratory results | | Numerical solution (friction) | |
|---|---|---|---|---|---|---|
| | | | Regime | $h_u$ (m) | Regime | $h_u$ (m) |
| L1 | 0.170 | 0.565 | Non-orifice | - | Non-orifice | - |
| L2 | 0.180 | 0.533 | Non-orifice | - | Non-orifice | - |
| L3 | 0.185 | 0.519 | Non-orifice | - | Non-orifice | - |
| L4 | 0.190 | 0.505 | $t^* = 0.5$ s | $0.105^*$ | $t^* = 0.1$ s | $0.097^*$ |
| L5 | 0.195 | 0.492 | $t^* = 2$ s | $0.110^*$ | $t^* = 1.8$ s | $0.108^*$ |
| L6 | 0.20 | 0.480 | Orifice | 0.130 | Orifice | 0.118 |

Table 3. Laboratory dam-break tests: initial conditions, exact frictionless solution, numerical frictionless solution.

| Test | $h_L$ (m) | $a/h_L$ | Exact solution | | Numerical solution (no friction) | |
|---|---|---|---|---|---|---|
| | | | Regime | $h_u$ (m) | Regime | $h_u$ (m) |
| L1 | 0.170 | 0.565 | Non-orifice | - | Non-orifice | - |
| L2 | 0.180 | 0.533 | Non-orifice | - | Non-orifice | - |
| L3 | 0.185 | 0.519 | Non-orifice | - | Non-orifice | - |
| L4 | 0.190 | 0.505 | Non-orifice | - | Non-orifice | - |
| L5 | 0.195 | 0.492 | Orifice | 0.110 | Orifice | 0.110 |
| L6 | 0.20 | 0.480 | Orifice | 0.119 | Orifice | 0.119 |

Figure 1. Relationship between flow and sluice gate: orifice free flow conditions (a); orifice submerged flow conditions (b); non-orifice flow regime (c).

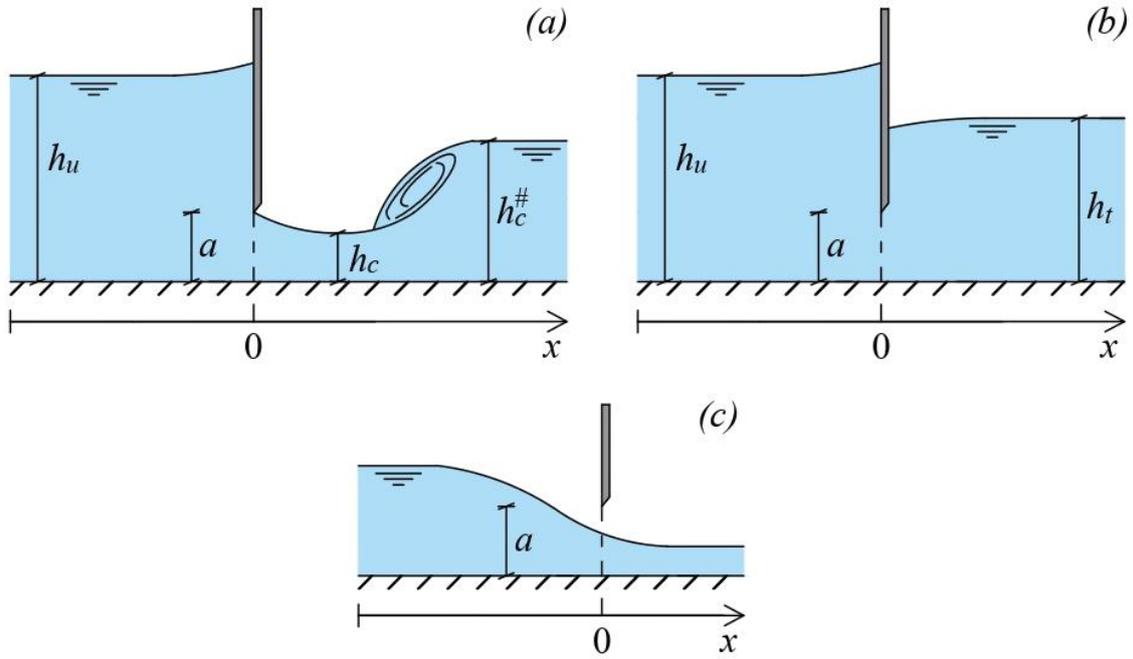

Figure 2. Contraction coefficient $C_c$ (a) and squared Froude number $F_F^2$ (b) as functions of the relative opening $a/h_u$.

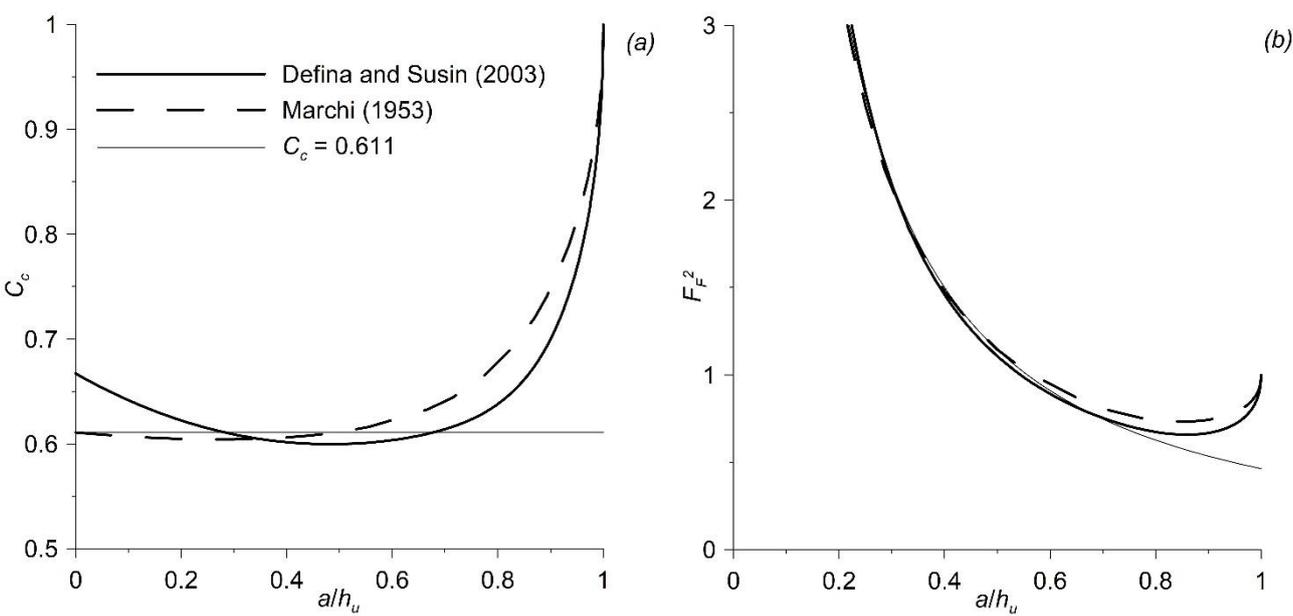

Figure 3. Dam-break on dry bed with $a = 0.47$ m and $h_L = 1$ m (Test E1 of Table 1), exact solutions (flow depth) at $t = 5$ s: non-orifice solution (a); free flow solution with $\mathbf{u}_1 = \mathbf{u}_{F,l}$ (b); free flow solution with $\mathbf{u}_1 = \mathbf{u}_{F,h}$ (c).

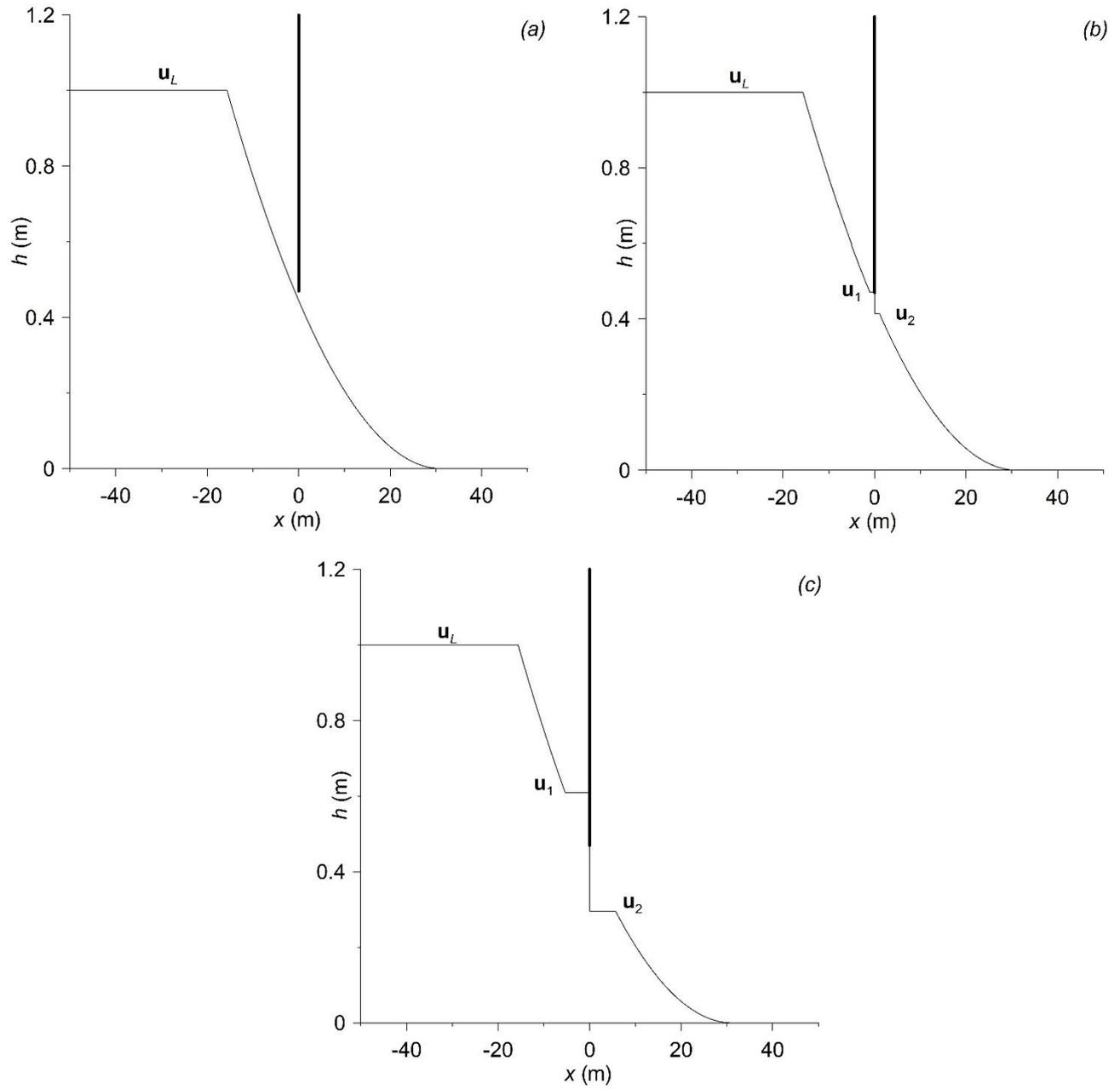

Figure 4. Dam-break on dry bed: relative flow depth $h_1/h_L$ as a function of the initial relative opening $a/h_L$.

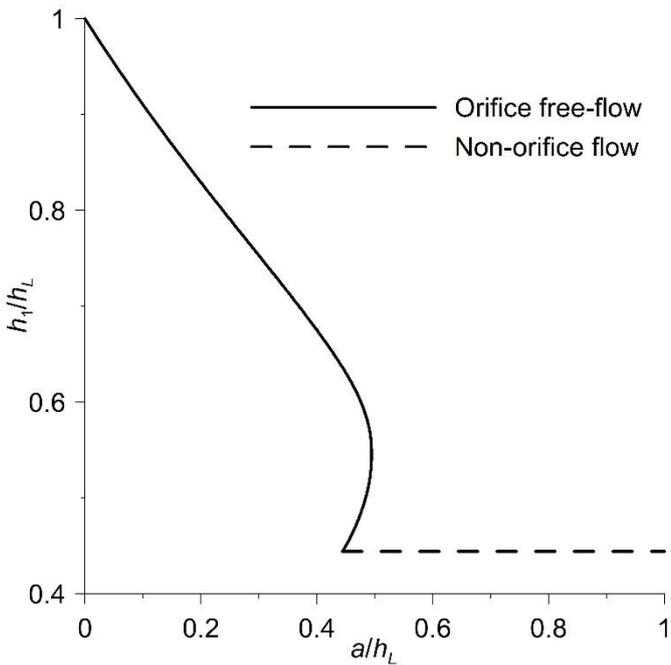

Figure 5. Exact solutions (flow depth) at $t = 5$ s for the dam-break tests of Table 1 with $a/h_L = 0.2$: Test E2 (a); Test E3 (b); Test E4 (c).

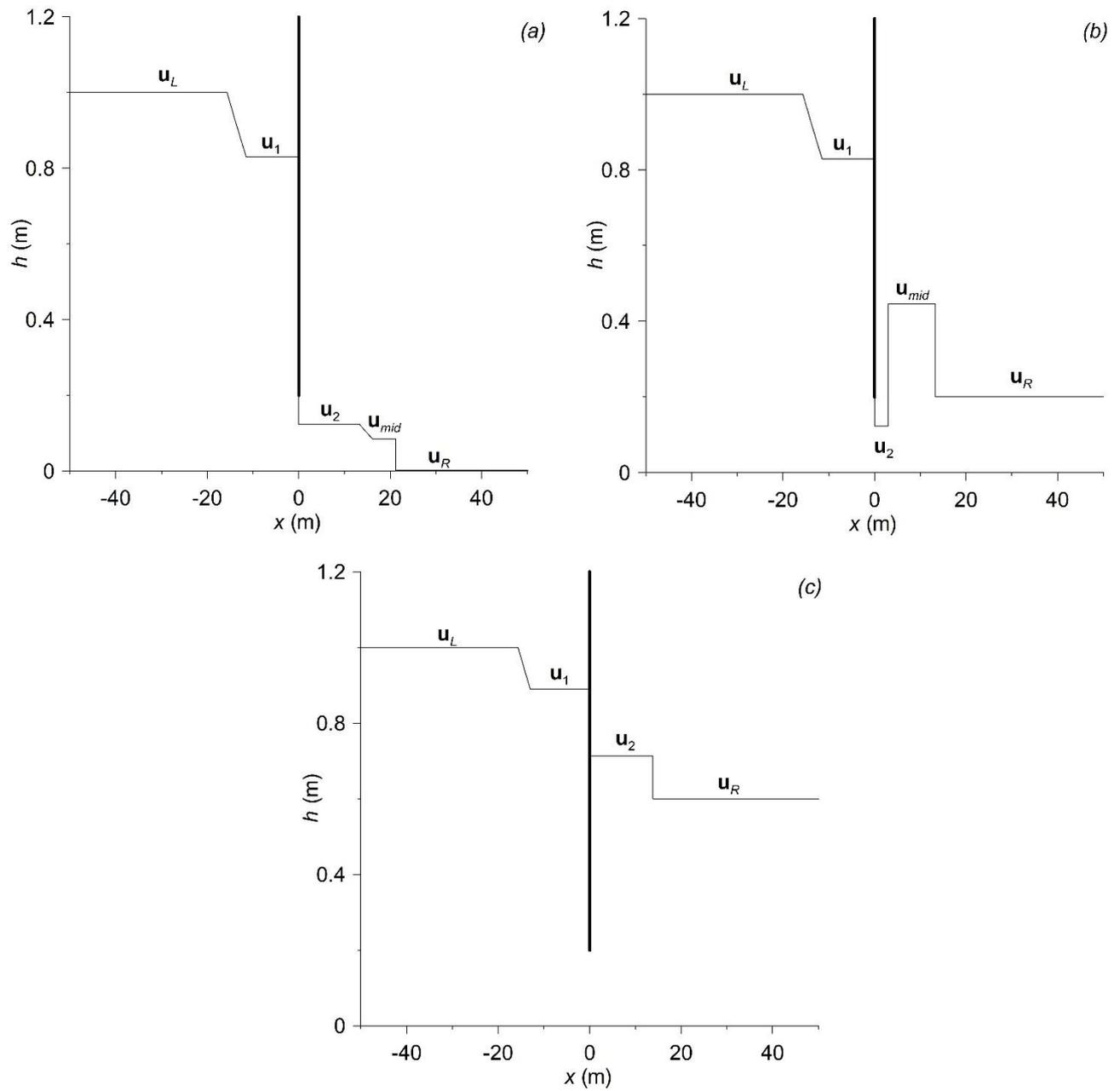

Figure 6. Construction of the exact solution for the dam-break with $a/h_L = 0.2$: L-M curve (a); solution of Test E2 (b); solution of Test E3 (c); solution of Test E4 (d).

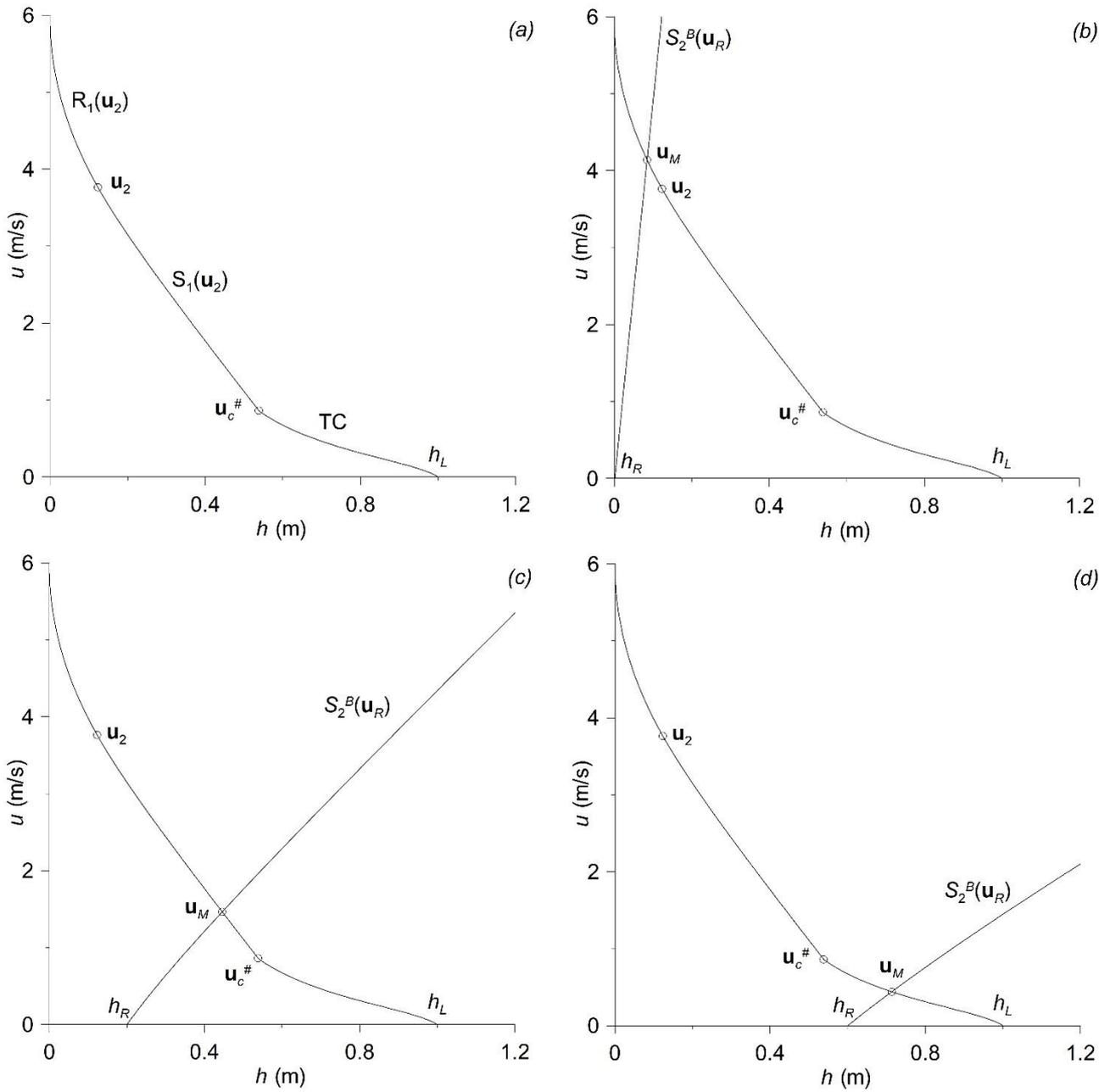

Figure 7. Exact solutions (flow depth) at $t = 5$ s for the dam-break tests of Table 1 with $a/h_L = 0.6$: Test E5 (a); Test E6 (b).

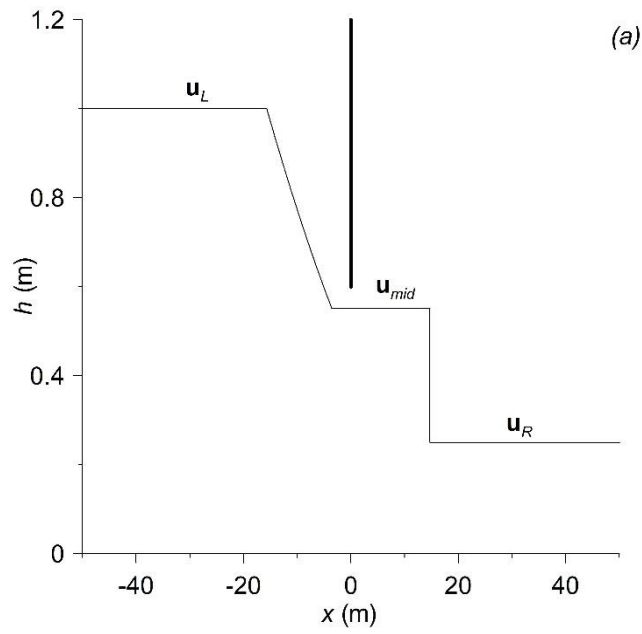
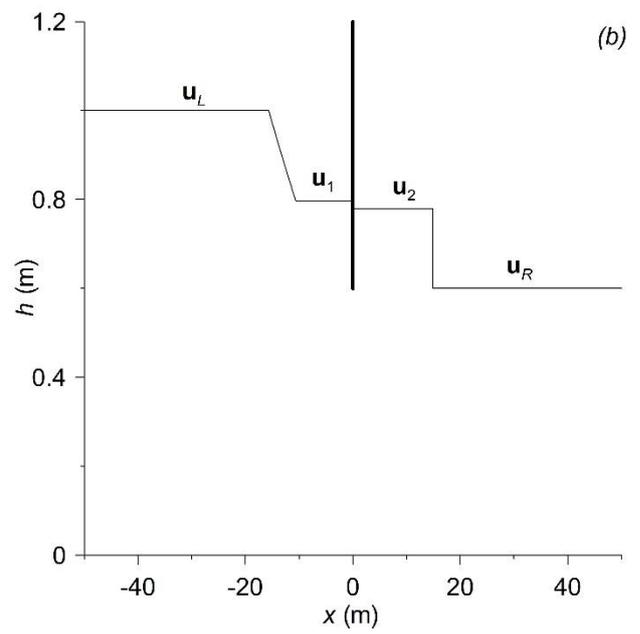

Figure 8. Construction of the exact solution for the dam-break with $a/h_L = 0.6$: L-M curve (a); solution of Test E5 (b); solution of Test E6 (c).

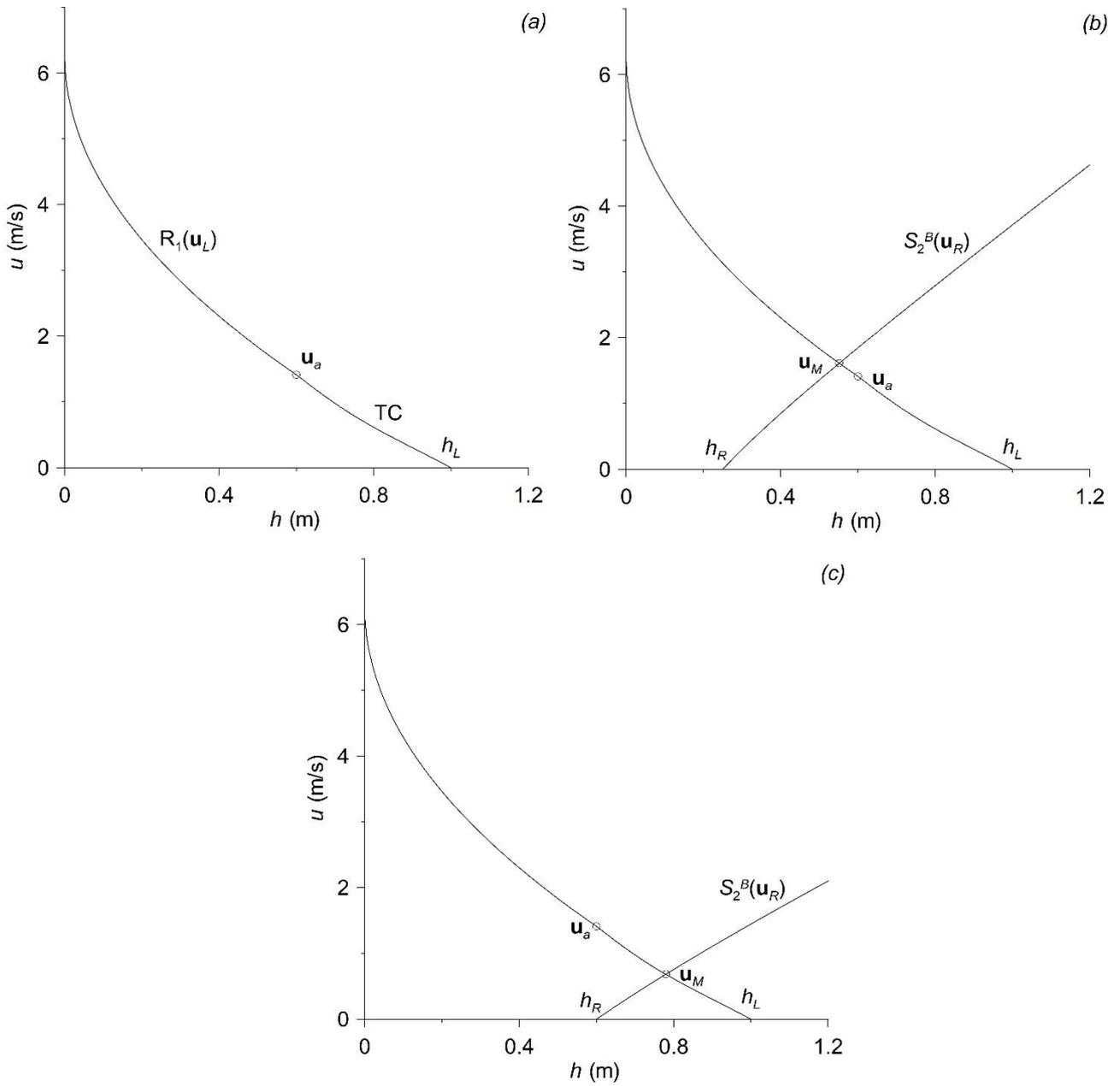

Figure 9. Exact solutions (flow depth) connected to $\mathbf{u}_{F,h}$ at $t = 5$ s for the dam-break tests of Table 1 with $a/h_L = 0.47$: Test E7 (a); Test E8 (b); Test E9 (c).

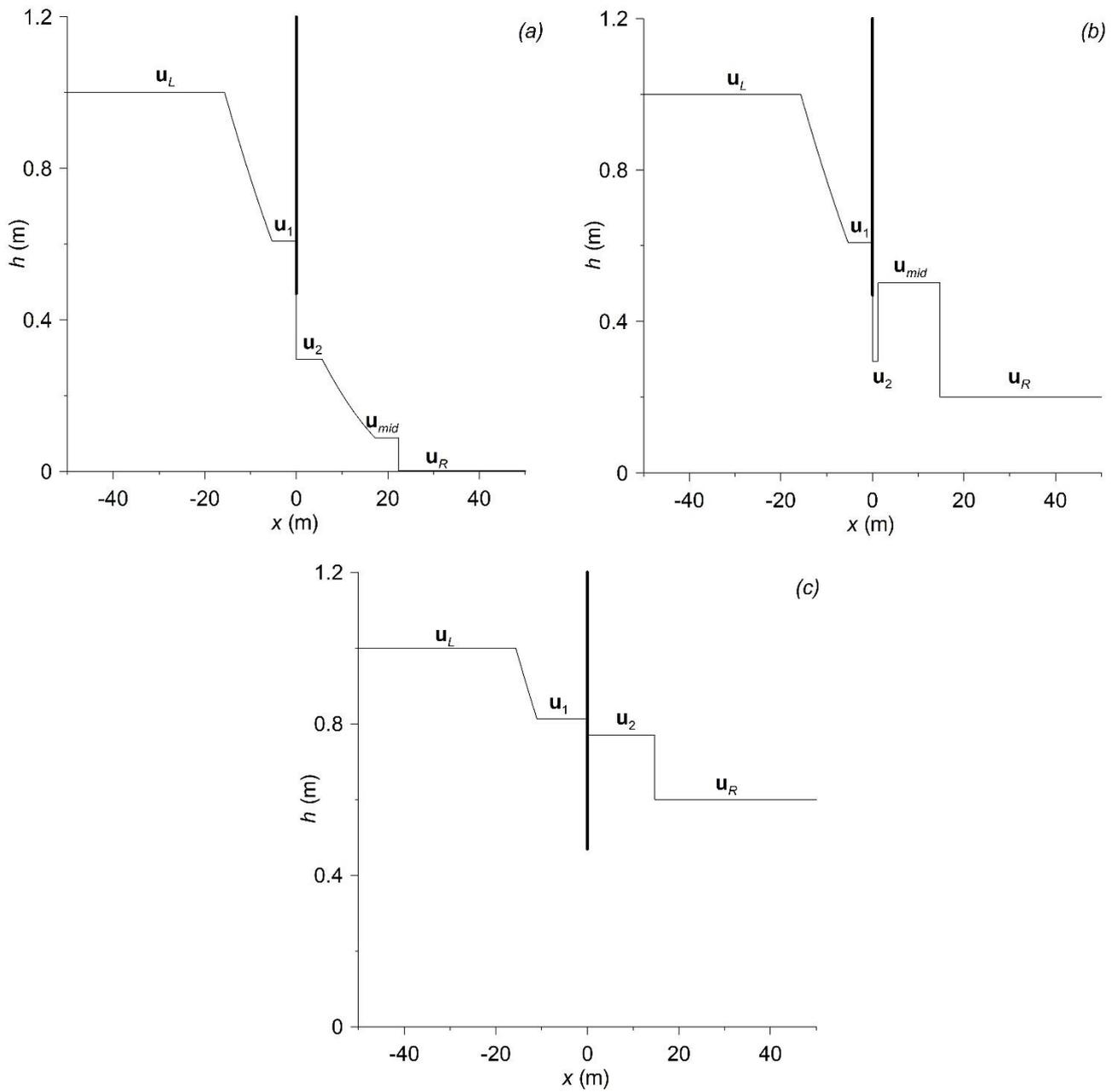

Figure 10. Construction of the exact solutions for the dam-break with $a/h_L = 0.47$: L-M curve connected to $\mathbf{u}_{F,h}$ (a); solution of Test E7 (b); solution of Test E8 (c); solution of Test E9 (d).

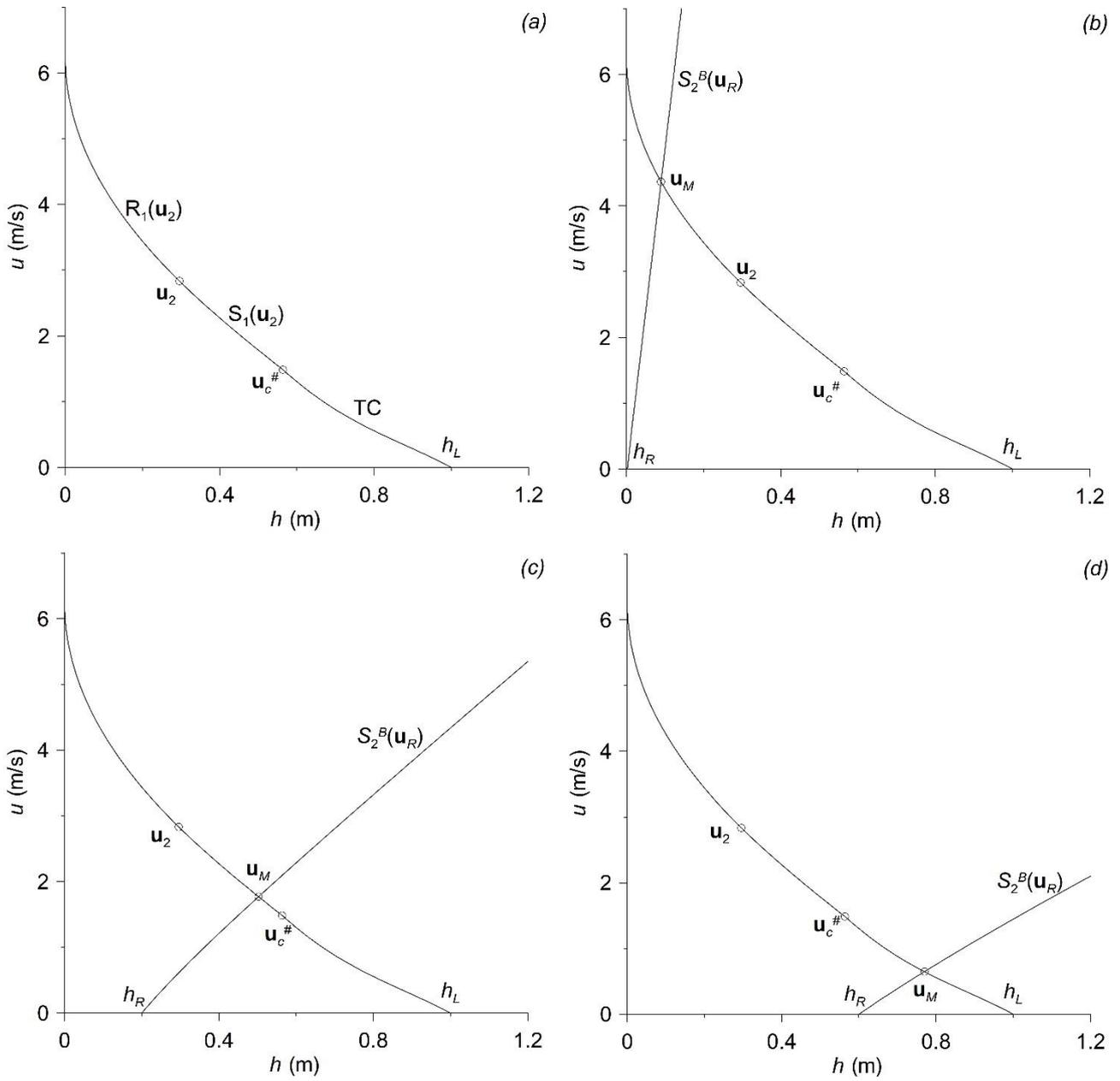

Figure 11. L-M curves with a gap for the dam-break with $a/h_L = 0.47$: L-M curve related to $\mathbf{u}_{F,l}$ (a); L-M curve related to the non-orifice solution of the dam-break on dry bed (b).

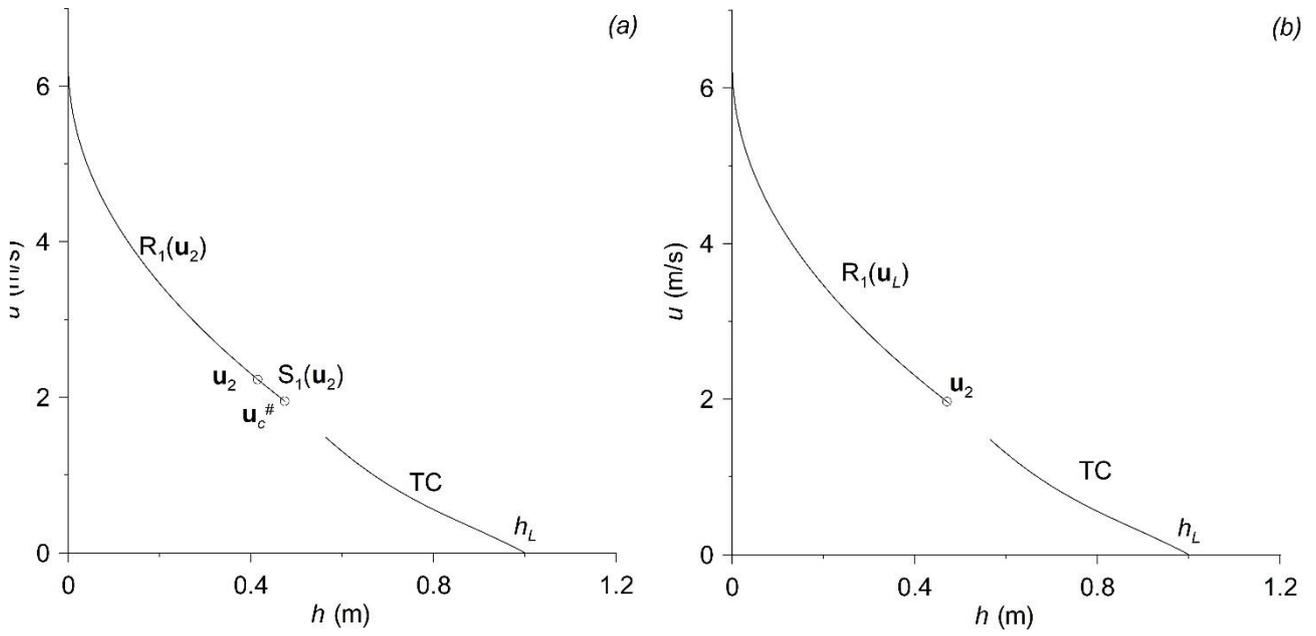

Figure 12. Unit-width discharges $q_{R,1}$ and $q_F$ for the construction of the TC curve in the case $h_L = 1$ m and $a = 0.47$ m: L-M curve related to $\mathbf{u}_{F,h}$ (a); L-M curve related to $\mathbf{u}_{F,l}$ (b); L-M curve related to the non-orifice solution of the dam-break on-dry bed (c).

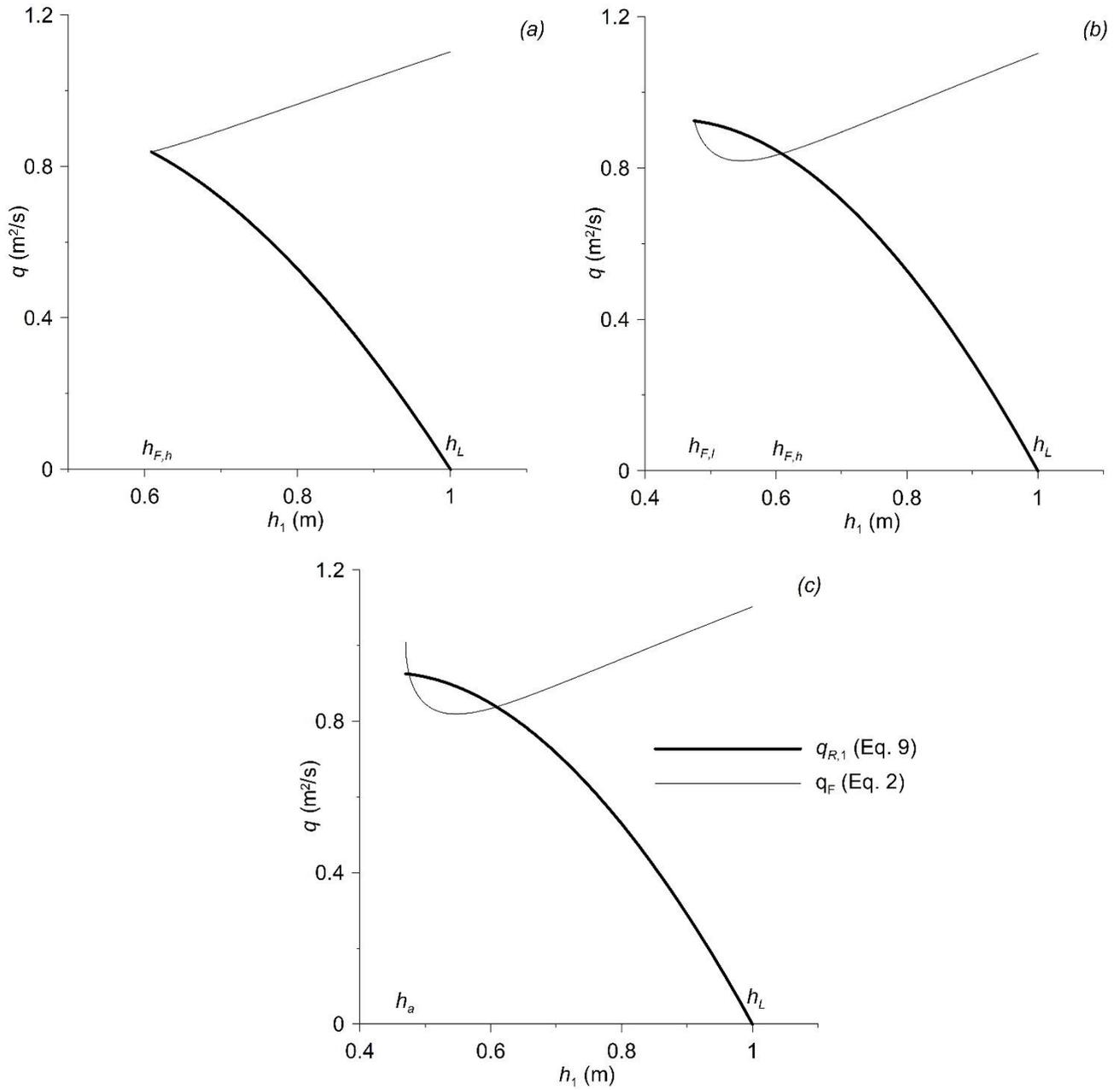

Figure 13. Dam-break on dry bed with *a* = 0.47 m and $h_L$ = 1 m (Test E1 of Table 1). Flow depth at *t* = 5 s: exact solution (thin black line); numerical solution with the equilibrium approach (dots).

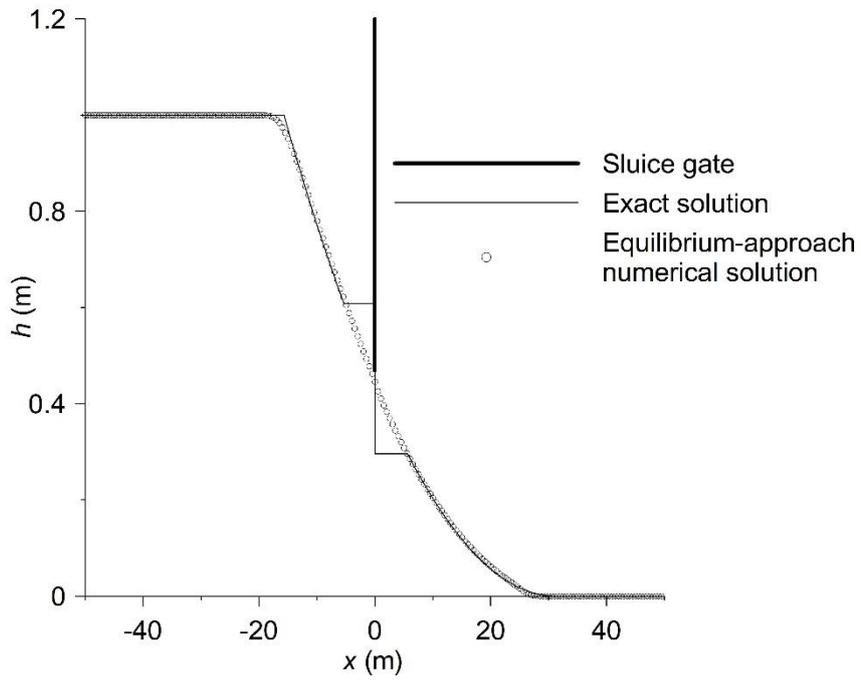

Figure 14. Dam-break on dry bed with $a = 0.47$ m and $h_L = 1$ m (Test E1 of Table 1) with the equilibrium approach. Time graphs of unit-width discharge under the gate (a); flow depth upstream of the gate (b); relative opening (c).

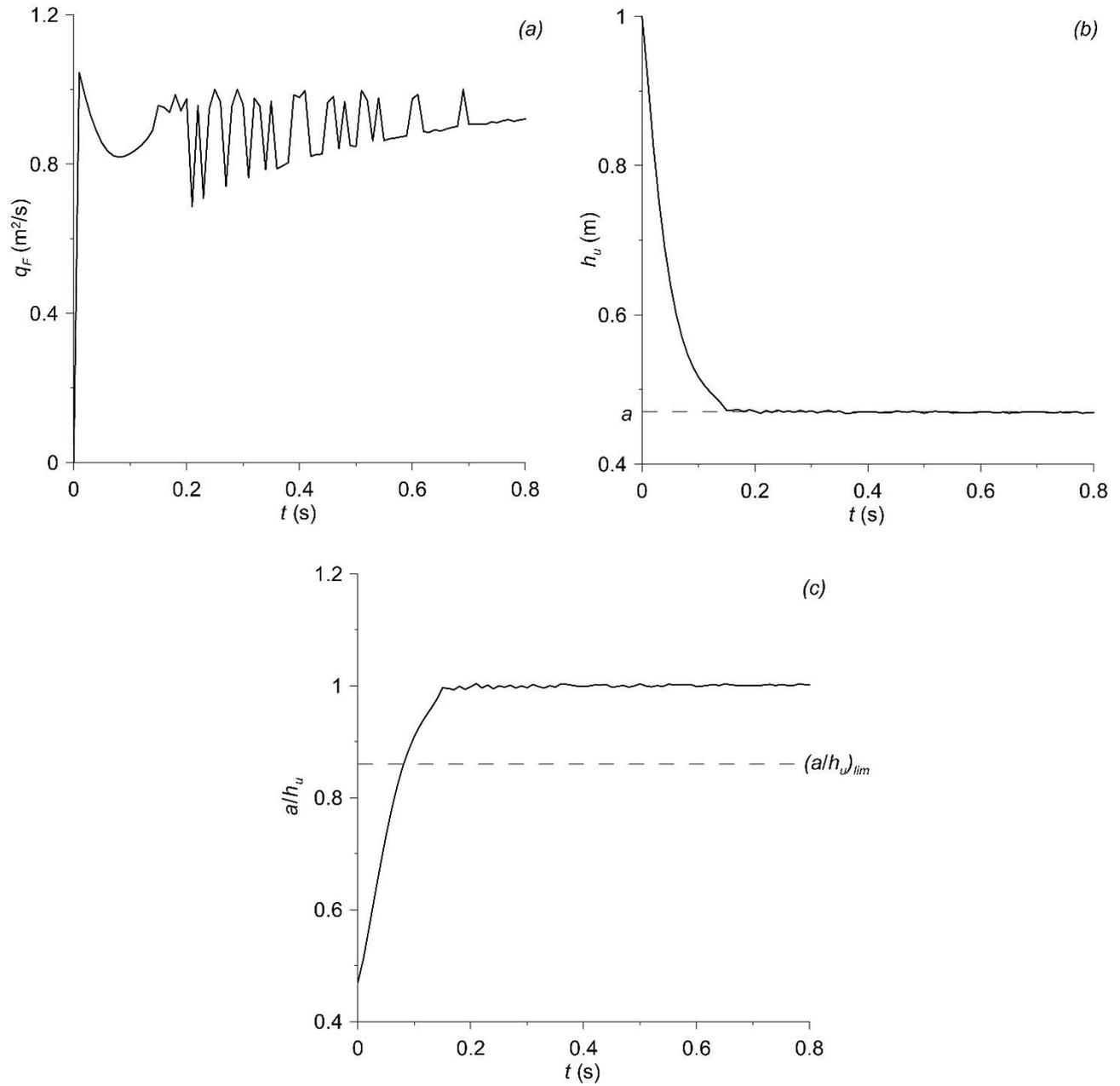

Figure 15. Dam-break on dry bed with $a = 0.47$ m and $h_L = 1$ m (Test E1 of Table 1). Flow depth at $t = 5$ s: exact solution (thin black line); numerical solution with the non-equilibrium approach (dots).

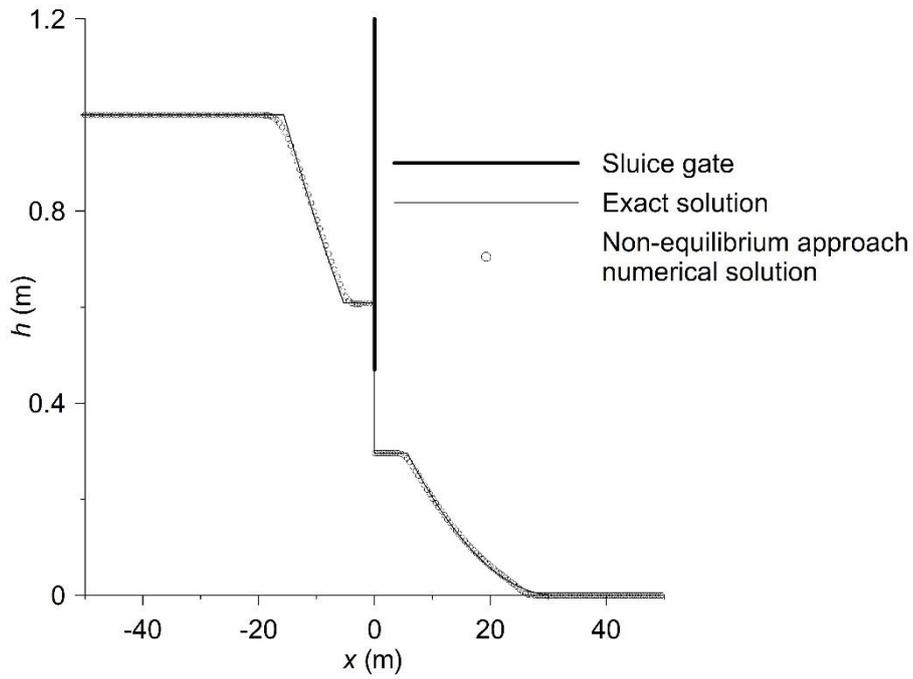

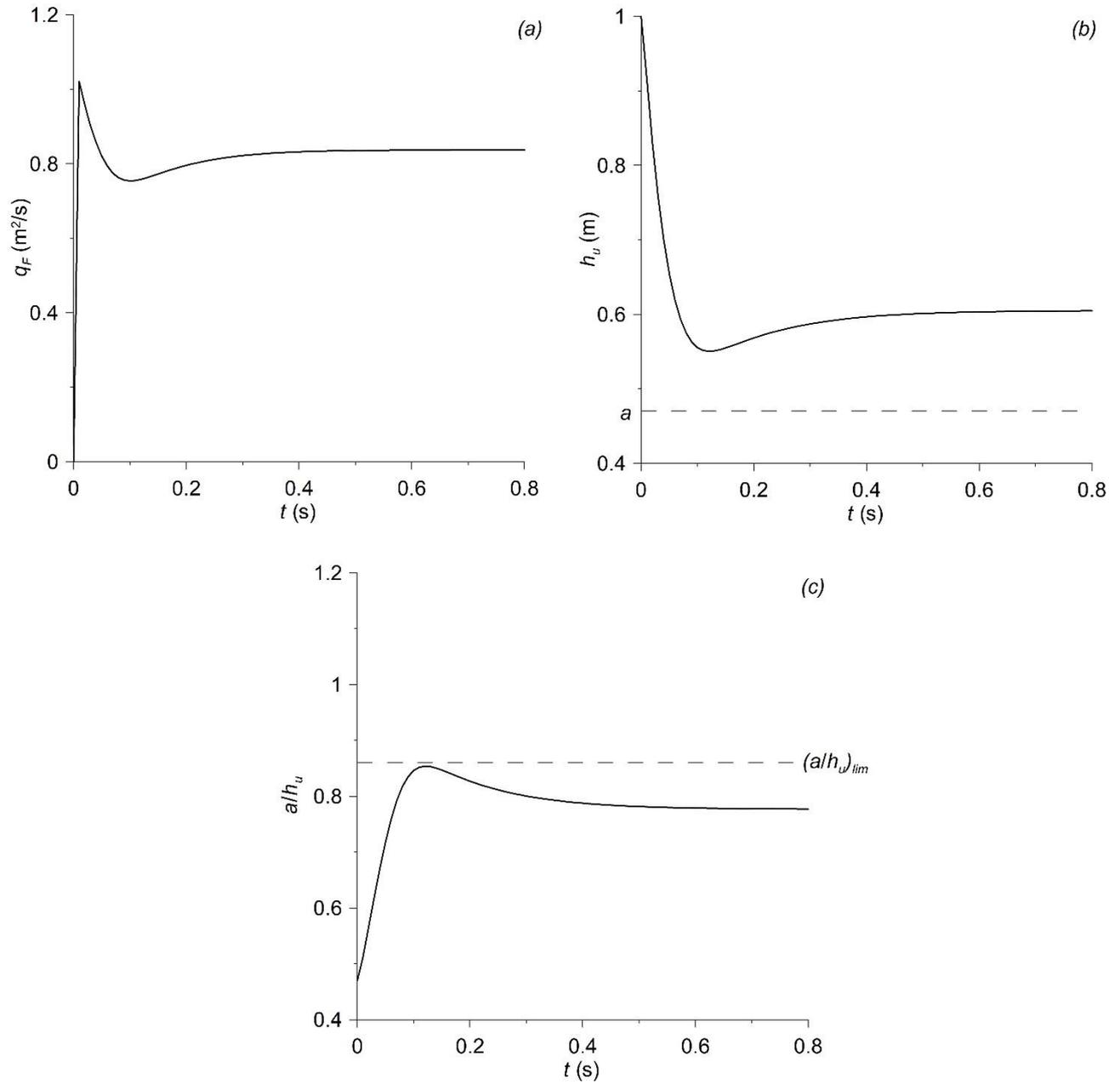

Figure 16. Dam-break on dry bed with $a = 0.47$ m and $h_L = 1$ m (Test E1 of Table 1) with the non-equilibrium approach. Time graphs of unit-width discharge under the gate (a); flow depth upstream of the gate (b); relative opening (c).

Figure 17. Comparison between exact (thin black line) and numerical solution with the non-equilibrium approach (dots, one in five is represented to enhance the clarity of the plot) for the dam-break tests of Table 1 with $a/h_L = 0.2$. Flow depth at $t = 5$ s: Test E2 (a); Test E3 (b); Test E4 (c).

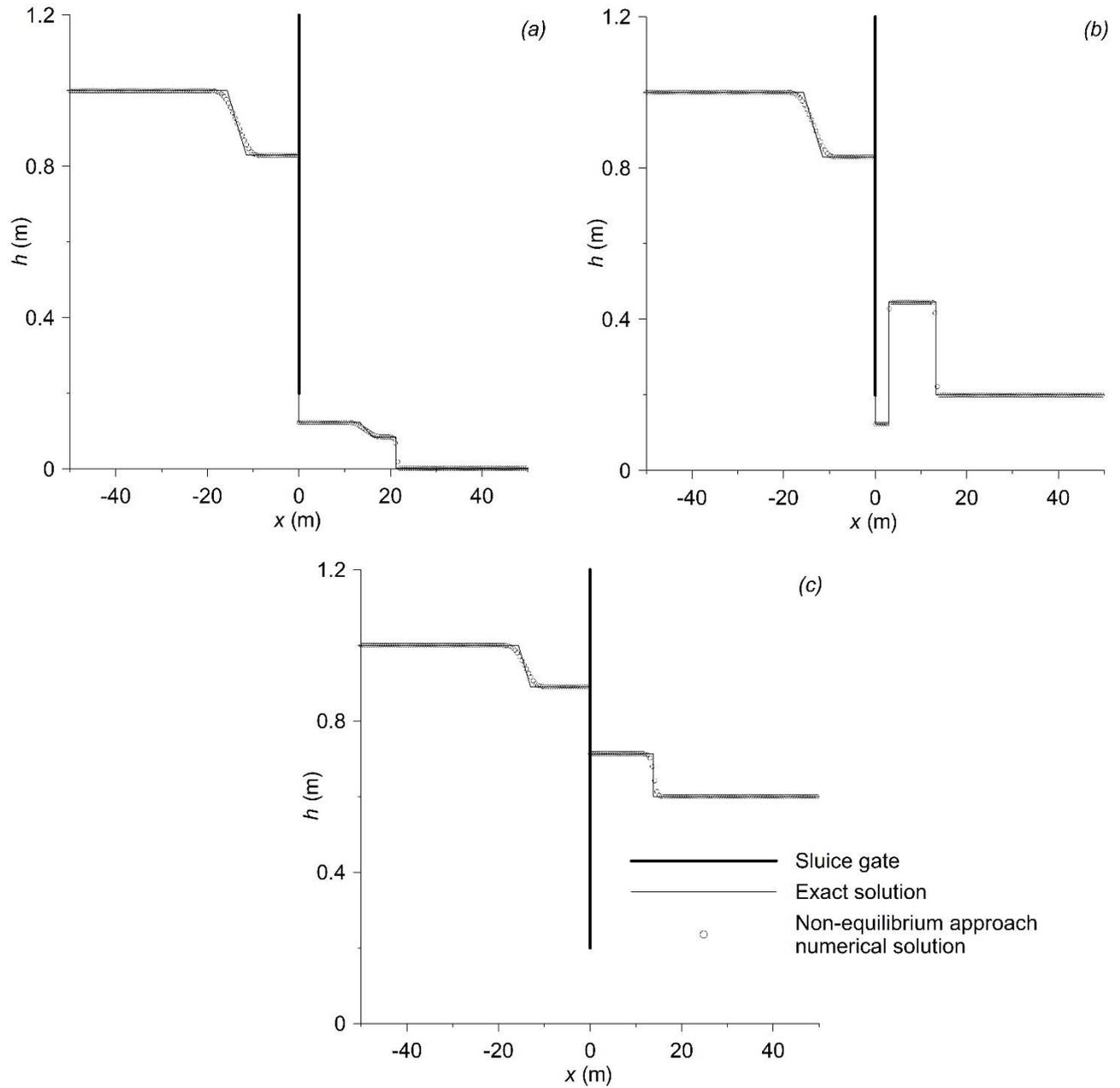

Figure 18. Comparison between exact (thin black line) and numerical solution with the non-equilibrium approach (dots, one in five is represented to enhance the clarity of the plot) for the dam-break tests of Table 1 with $a/h_L = 0.6$. Flow depth at $t = 5$ s: Test E5 (a); Test E6 (b).

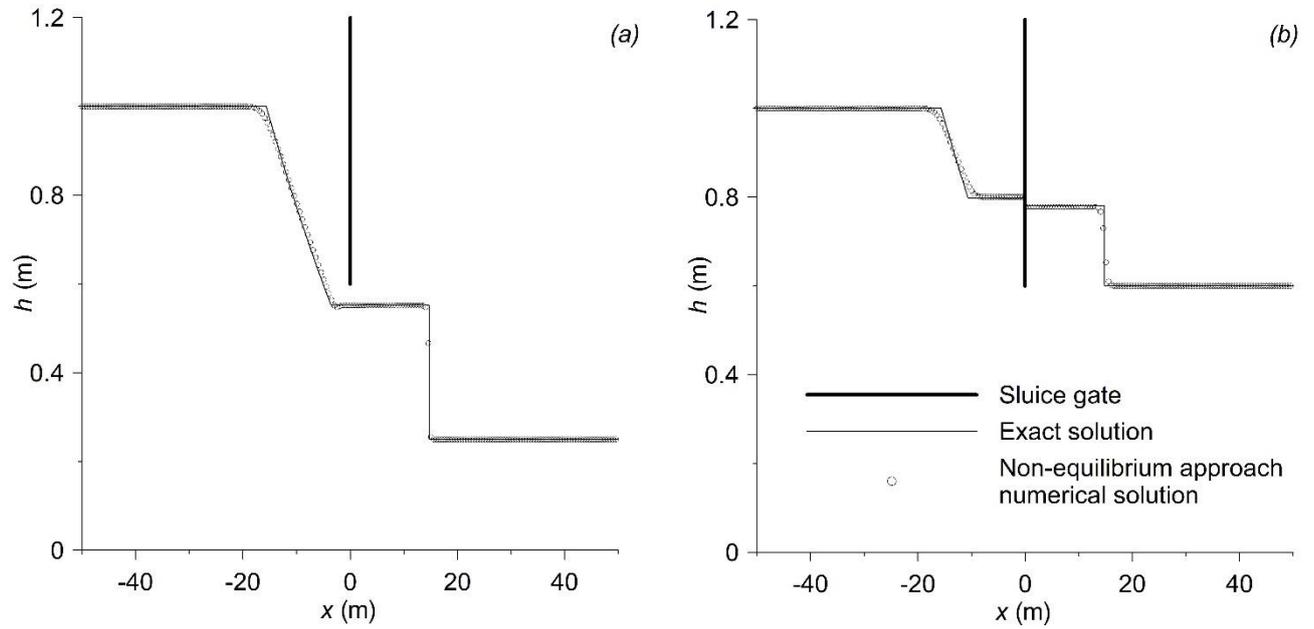

Figure 19. Comparison between exact (thin black line) and numerical solution with the non-equilibrium approach (dots, one in five is represented to enhance the clarity of the plot) for the dam-break tests of Table 1 with $a/h_L = 0.47$. Flow depth at $t = 5$ s: Test E7 (a); Test E8 (b); Test E9 (c).

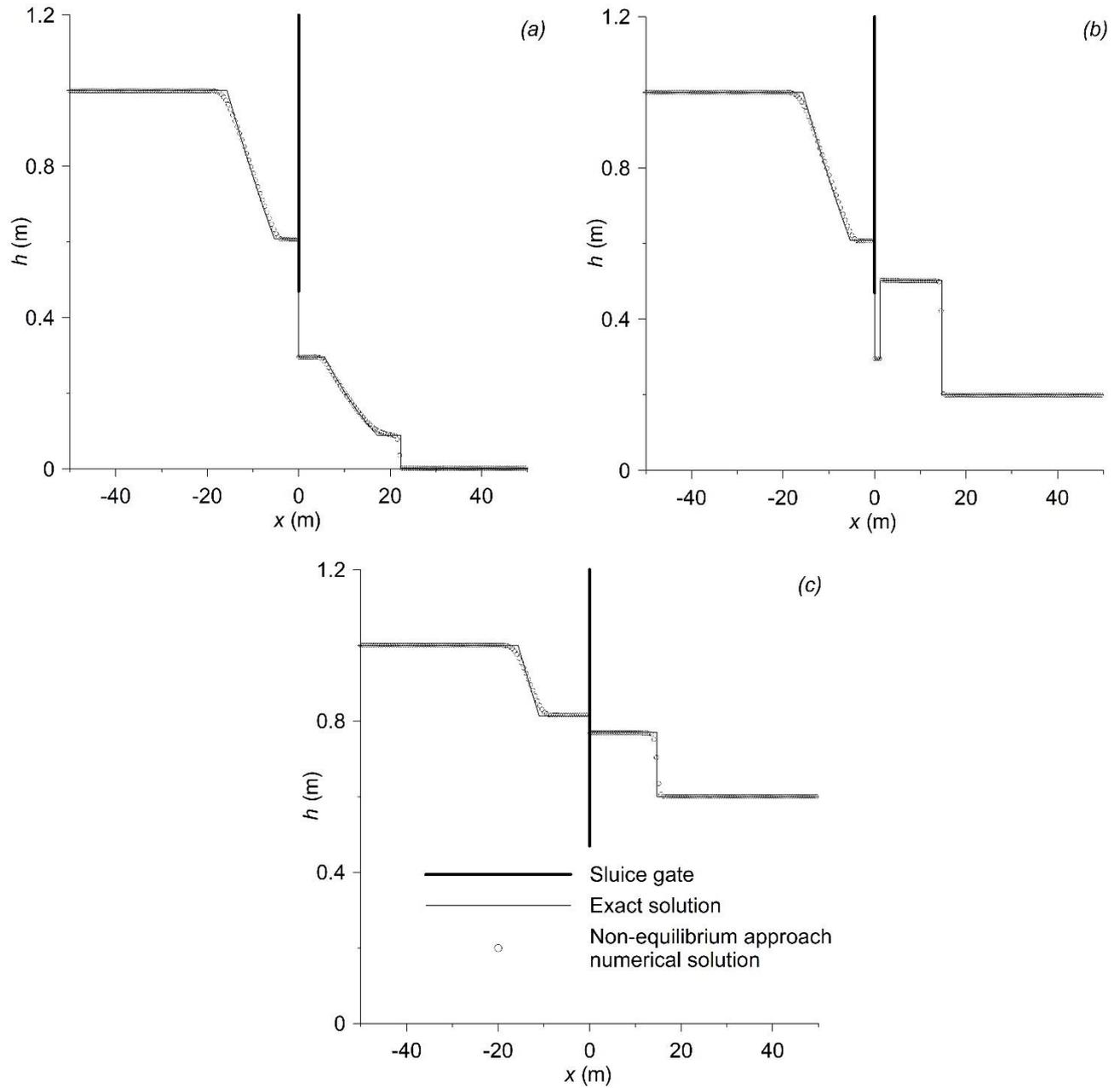

Figure 20. Numerical time histories (thin black line) of the upstream flow depth for the dam-break experiments of Table 2: Test L1 (a); Test L2 (b); Test L3 (c); Test L4 (d); Test L5 (e); Test L6 (f). The position of the gate lip is represented with a dashed line.

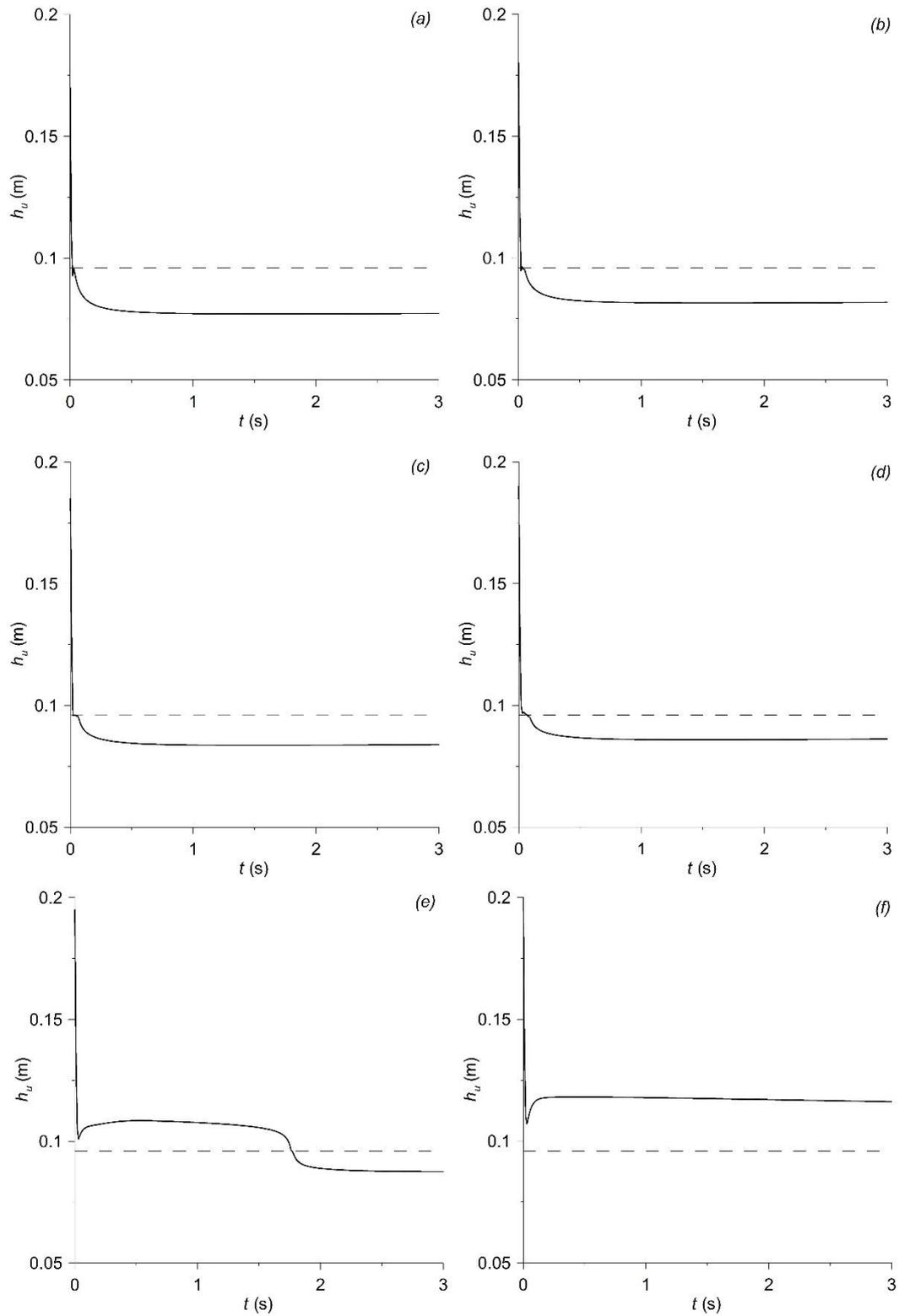

Figure 21. Sluice gate plan-view: fixed reference $Oxy$ (a); reference $O'xy'$ translating with uniform velocity $v_u$ (b).

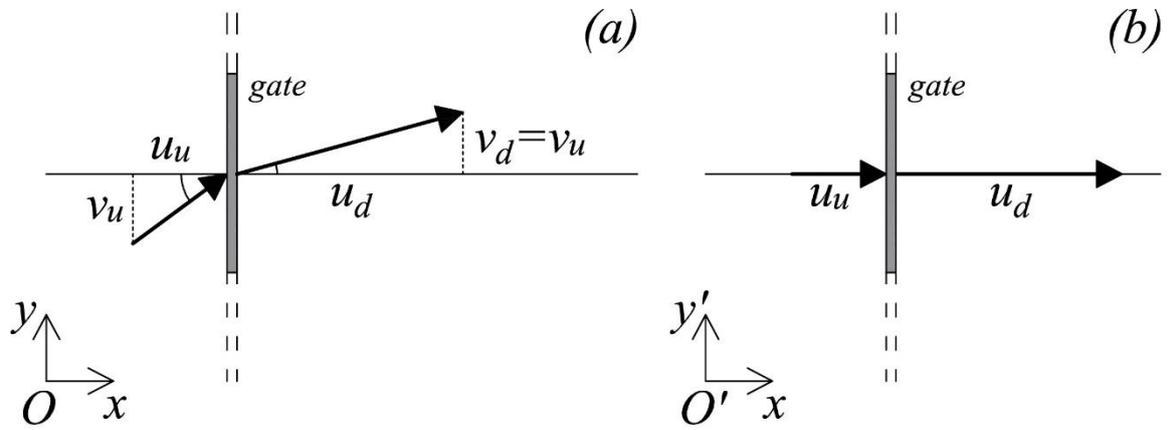

Figure 22. Geometry of the detention basin: plan view (a); longitudinal section A-A (b); transverse cross-section B-B (c). Distorted representation. Measures in metres.

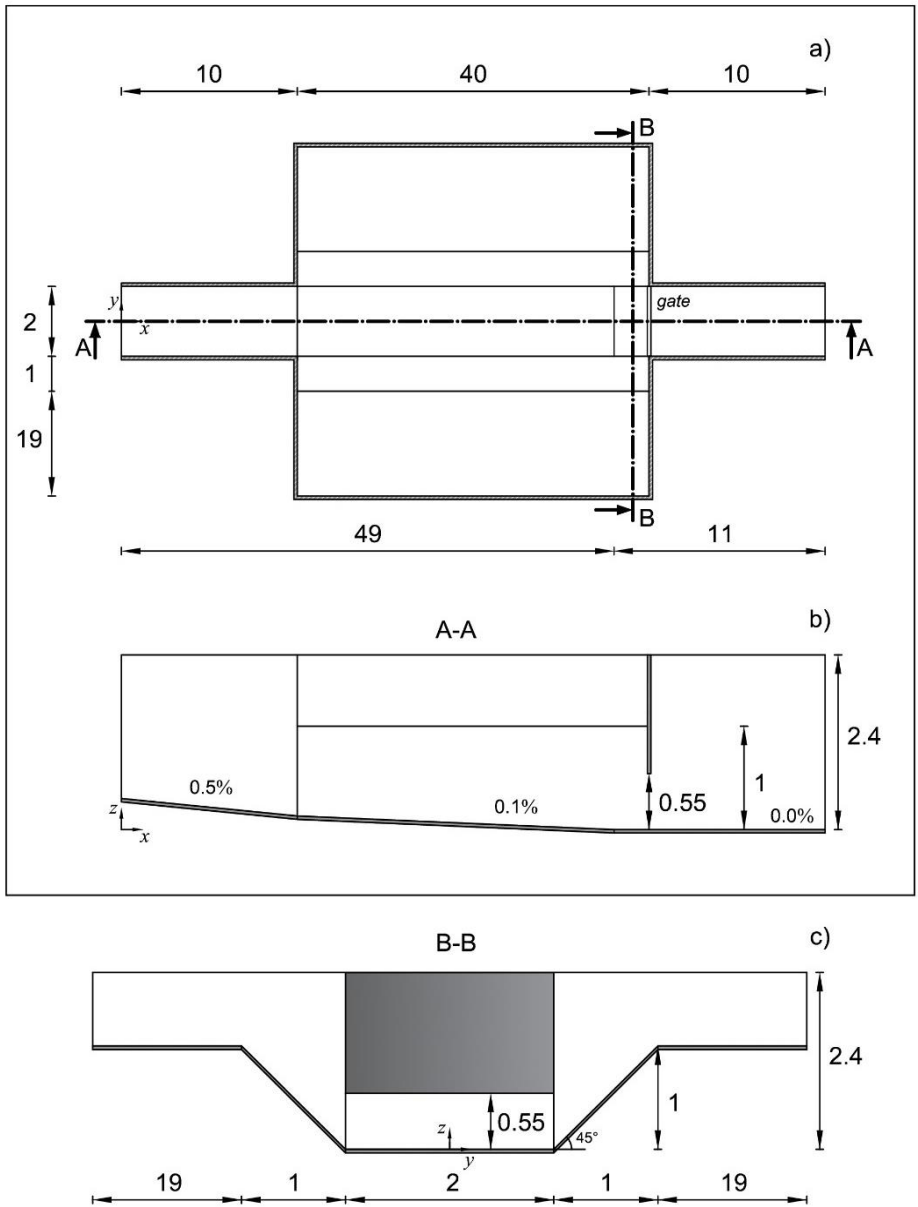

Figure 23. Inflow to the detention basin: discharge (thin black line) and flow depth (thick black line) hydrographs.

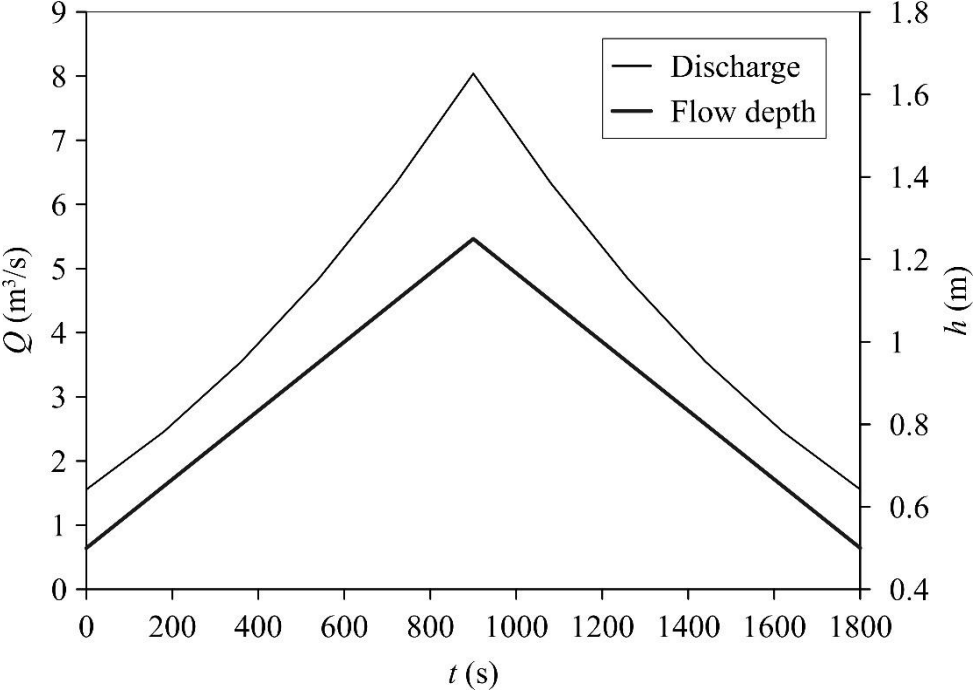

Figure 24. Free-surface profile along the detention basin longitudinal axis at different times: $t = 0$ s (a); $t = 300$ s (b), $t = 1320$ s (c), $t = 1800$ s (d).

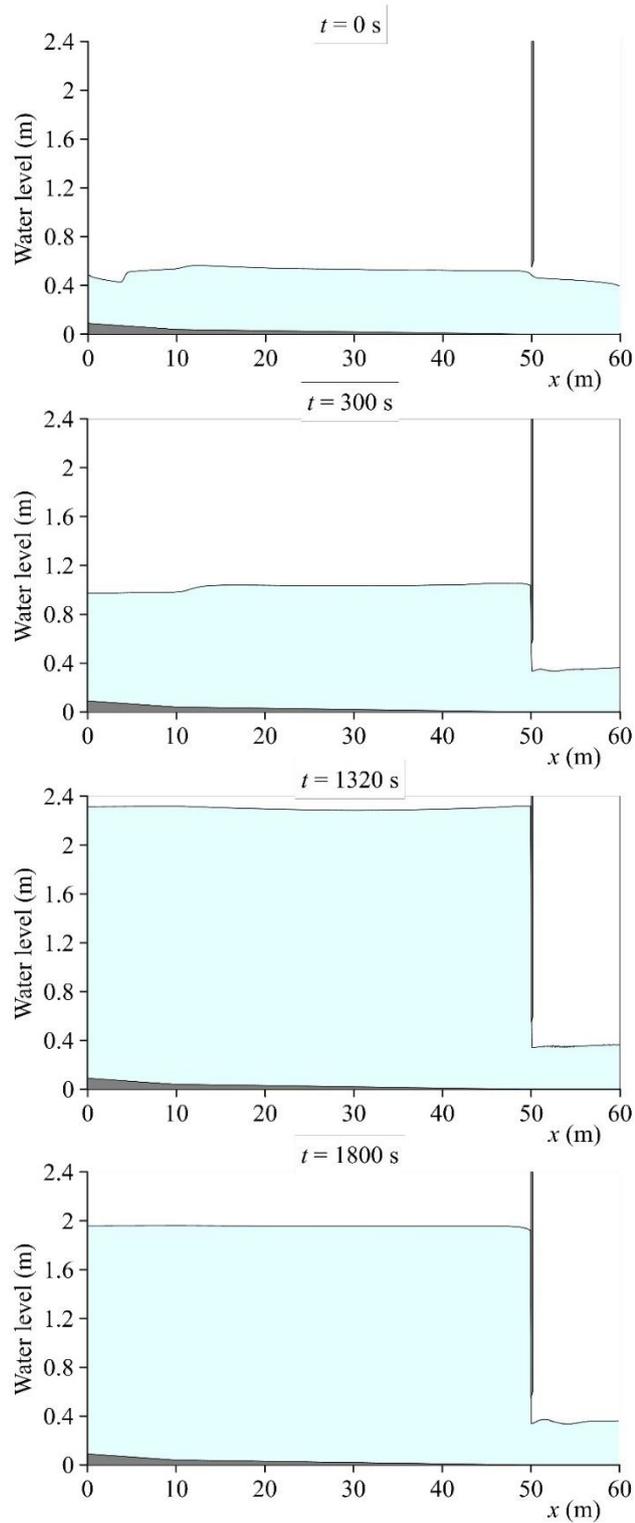

Figure 25. Detention basin inflow and outflow discharges: inflow discharge (thin black line); outflow discharge with gate (thick black line); outflow discharge without gate (dashed line). The arrows individuate the instants corresponding to times $t = 0$ s (a), $t = 300$ s (b), $t = 1320$ s (c), $t = 1800$ s (d), respectively.

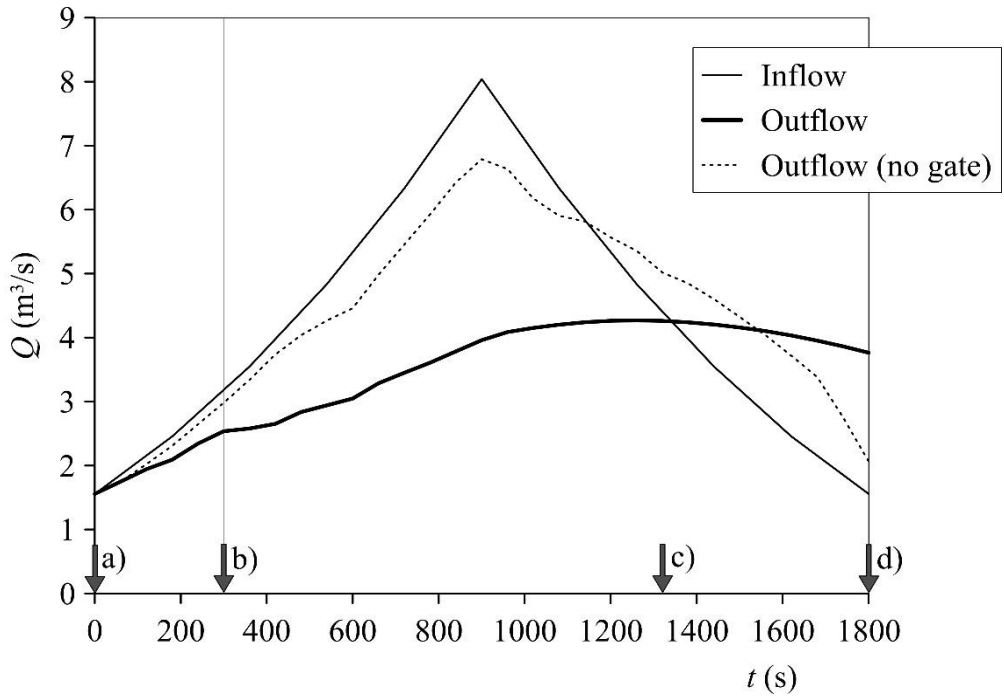

Figure 26. Admissible solutions of the dam-break on dry bed with partially lifted sluice gate: disambiguation criterion of Section 3.3 (a); disambiguation criterion by Lazzarin et al. (2023).

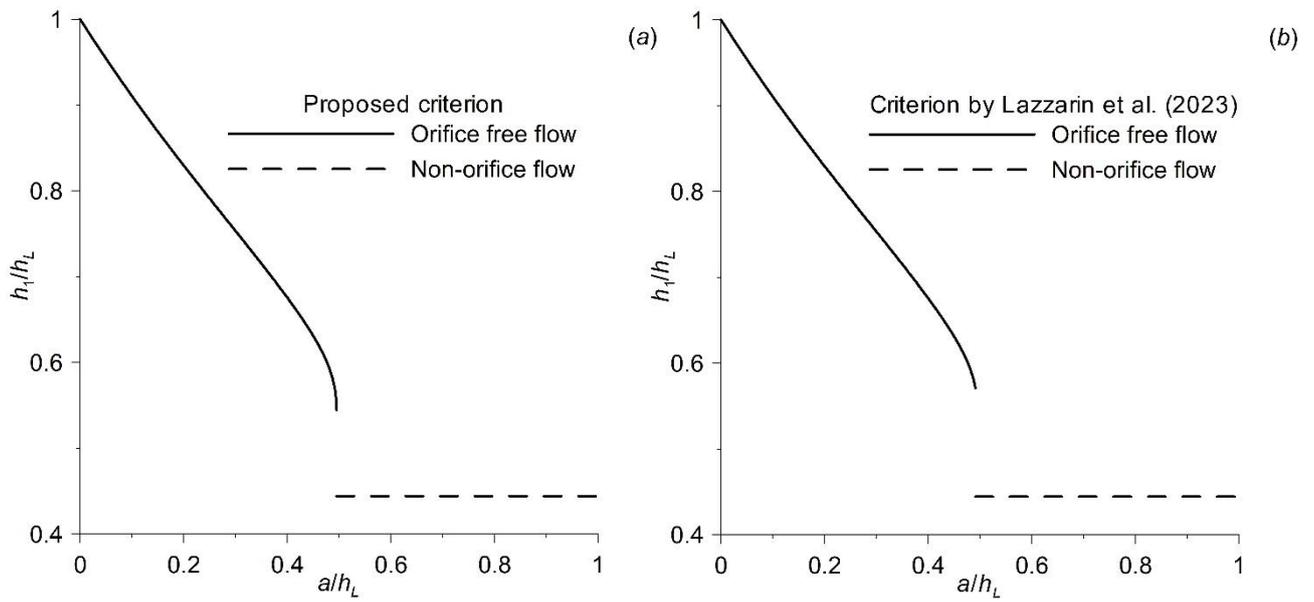

Figure 27. Numerical solution of the L5 dam-break problem (Table 2) with the numerical model of Section 4.2 (thin black line). Flow depth immediately upstream the gate: simulation without friction (a); simulation with friction (b). The position of the gate lip is represented with a dashed line.

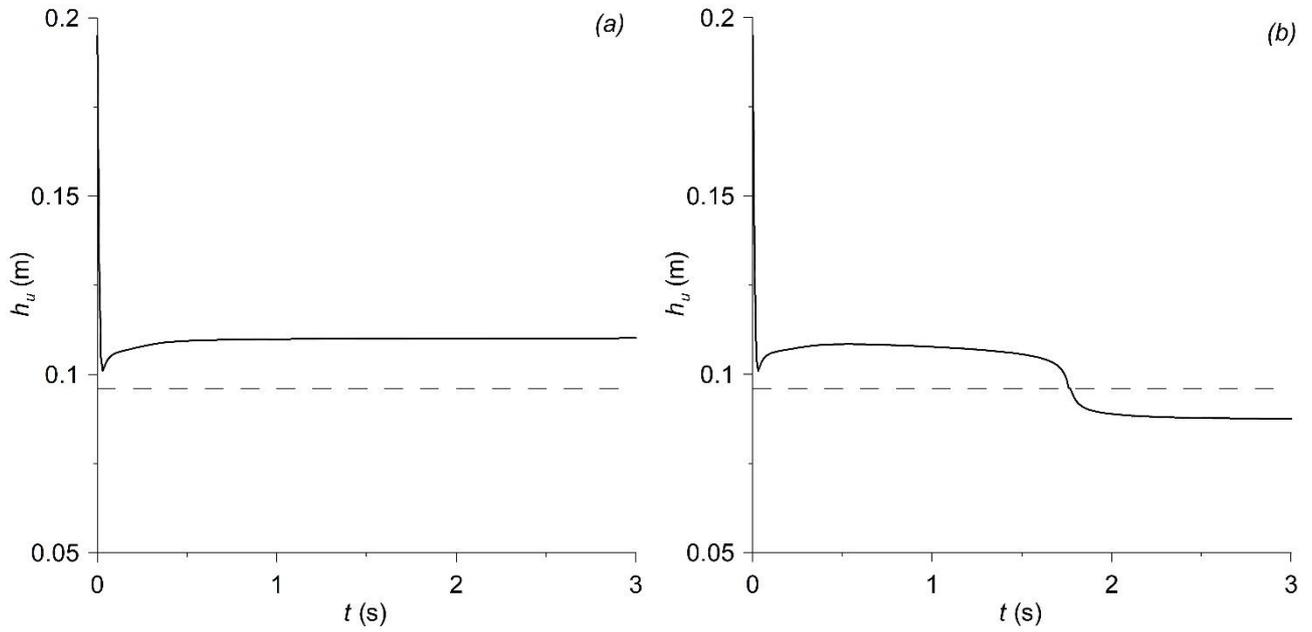

Figure 28. Numerical solution of the L5 dam-break problem (Table 2) with the numerical model of Section 4.2. Flow depth along the flume at time $t = 1.7$ s: simulation without friction (a); simulation with friction (b).

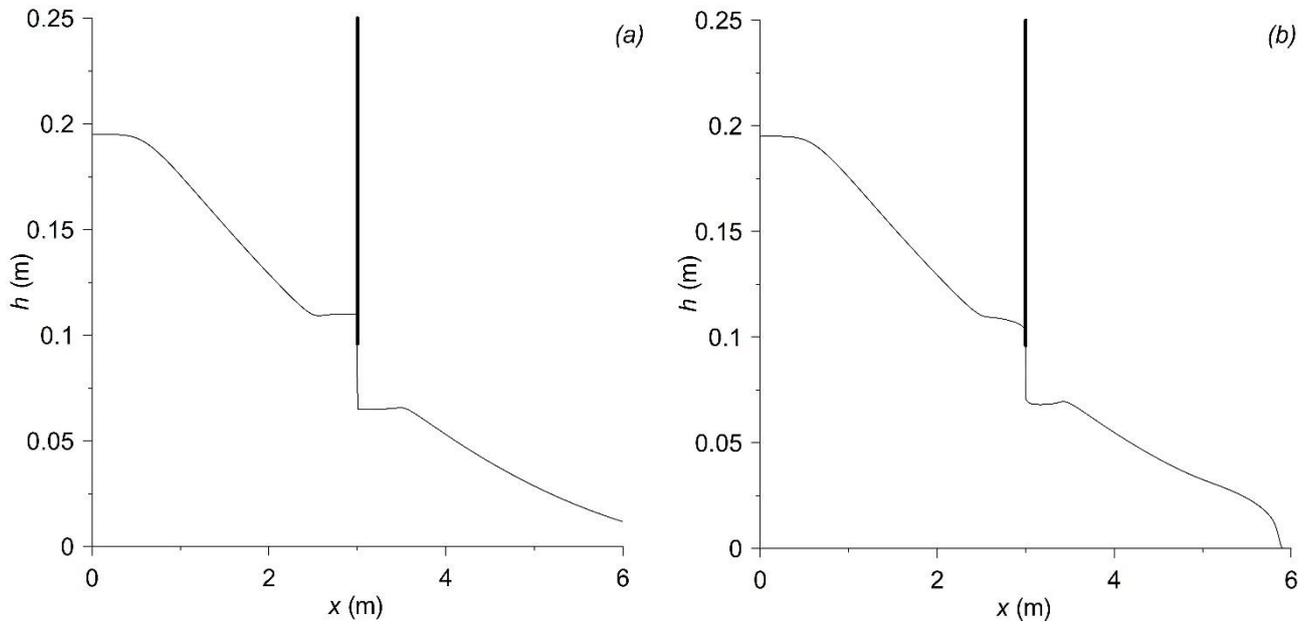

Figure 29. Physical justification of the non-equilibrium approach for the computation of the discharge issuing under the gate: particle trajectory under the gate from the upstream position A to the position B at the *vena contracta*.

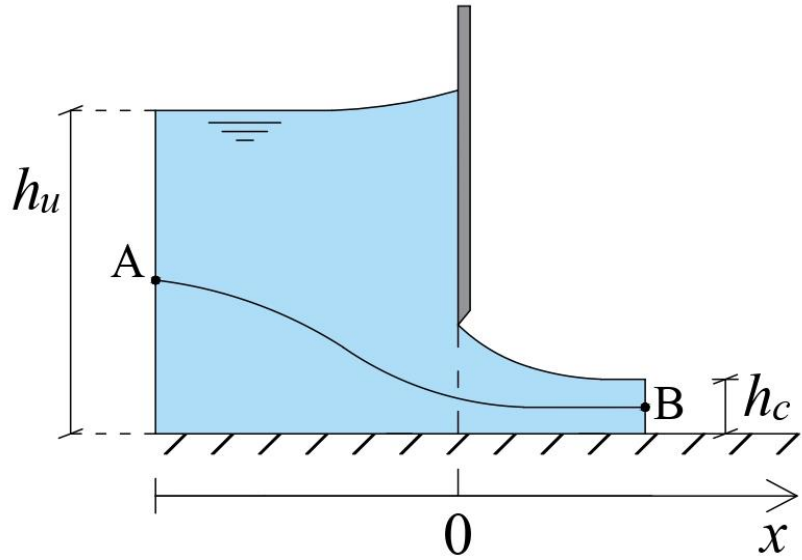